\documentclass[reprint, amsmath, amssymb, aps, showkeys]{revtex4-2}

\usepackage{gensymb}
\usepackage{booktabs}
\usepackage{graphicx}
\graphicspath{{./Figures/}}
\usepackage{multirow}
\usepackage{dcolumn}
\usepackage{bm}
\usepackage{times}
\usepackage{algorithm}
\usepackage{algorithmicx}
\usepackage{algpseudocode}
\usepackage{listings}
\usepackage{natbib}
\usepackage{placeins}
\usepackage{threeparttable}
\usepackage[table,dvipsnames]{xcolor}
\usepackage{longtable}
\usepackage{makecell}
\usepackage{array}
\usepackage{enumitem}
\usepackage[explicit]{titlesec}
\usepackage{xcolor}

\definecolor{RoyalBlue}{RGB}{65, 105, 225}

\usepackage[colorlinks = true,
linkcolor = blue,
urlcolor  = blue,
citecolor = blue,
anchorcolor = blue]{hyperref}

\newcommand{\beginsupplement}{%
  \setcounter{table}{0}
  \renewcommand{\thetable}{S\arabic{table}}%
  \setcounter{figure}{0}
  \renewcommand{\thefigure}{S\arabic{figure}}%
  \setcounter{section}{0}
  \renewcommand{\thesection}{\arabic{section}}
  \renewcommand{\thesubsection}{\thesection.\arabic{subsection}}
  \renewcommand{\thesubsubsection}{\thesubsection.\arabic{subsubsection}}
  \setcounter{equation}{0}
  \renewcommand{\theequation}{S\arabic{equation}}%
  \setcounter{page}{1}
}

\begin{document}
\newcolumntype{x}[1]{>{\centering\arraybackslash\hspace{0pt}}p{#1}}

\preprint{APS/123-QED}

\title{MatTools: Benchmarking Large Language Models for Materials Science Tools}

\author{Siyu Liu$^{1,2}$}
\author{Bo Hu$^{1}$}
\author{Beilin Ye$^{1}$}
\author{Jiamin Xu$^{1}$}
\author{David J. Srolovitz$^{1,2}$}\email{srol@hku.hk}
\author{Tongqi Wen$^{1,2}$}\email{tongqwen@hku.hk}

\affiliation{$^{1}$Center for Structural Materials, Department of Mechanical Engineering, The University of Hong Kong, Hong Kong SAR, China}
\affiliation{$^{2}$Materials Innovation Institute for Life Sciences and Energy (MILES), HKU-SIRI, Shenzhen, China}

\date{\today}

\begin{abstract}

The rapidly advancing capabilities of large language models (LLMs) are poised to transform materials science by automating key aspects of research. However, their proficiency in operating computational materials science tools remains largely unquantified. Here we introduce MatTools, the first dual-level multi-step reasoning benchmark designed to evaluate LLMs in this setting through question answering and real-world code execution. MatTools comprises (i) a QA benchmark of 69,225 question–answer pairs derived from the \texttt{pymatgen} (Python Materials Genomics) codebase and documentation, and (ii) a real-world tool-usage benchmark of 49 tasks (138 subtasks) requiring the generation and safe execution of Python code for materials property calculations. Evaluation across diverse LLMs yields three key insights: (1) \textbf{Generalists outshine specialists}: general-purpose LLMs significantly outperform materials science-focused LLMs in materials simulation tool knowledge assessments; (2) \textbf{AI knows AI}: documentation generated by LLMs outperforms both codebase and human-written documentation as retrieval sources; and (3) \textbf{Simpler is better}: LLMs utilizing only LLM-generated documentation as retrieval sources with self-reflection mechanisms outperform more complex multi-agent approaches such as Agentic RAG and GraphRAG. 
MatTools establishes as a standardized framework for systematically assessing and advancing LLM capabilities in computational materials science. It lays a foundation for developing more reliable, efficient, and scientifically effective AI systems, representing a critical step toward the integration of LLMs into next-generation scientific discovery.

\end{abstract}

\maketitle

\section{Introduction}
Large language models (LLMs) are now widely applied in materials science, with demonstrated utility in literature and materials database knowledge extraction~\citep{literature, Park2024-uo, Polak2024-sg}, materials property prediction~\citep{AtomGPT, das2023crysmmnet, liu2024llms}, alloy design~\citep{TIAN2025120663, alloyGPT}, discovering of physical laws~\citep{matlaws, du2024large}, and generation of scientific hypotheses~\citep{yang2024moose, alkan2025surveyhypothesisgenerationscientific}. 
An indispensable part of materials research lies in connecting experimental discoveries with mechanistic understanding through computational modeling. 
Moving beyond knowledge extraction and hypothesis generation, an emerging frontier is the use of LLMs to autonomously conduct such computational simulations/calculations. 
This represents a step change from the current paradigm, where researchers manually construct workflows for specific property evaluations. 
Realizing this capability requires LLMs to integrate competencies across domains: programming and algorithmic reasoning from computer science, workflow and code design from computational materials science, domain knowledge from materials physics and chemistry, and the orchestration of diverse computational tools to solve complex scientific problems.

\begin{figure*}[htbp]
	\centering
	\includegraphics[width=0.95\textwidth]{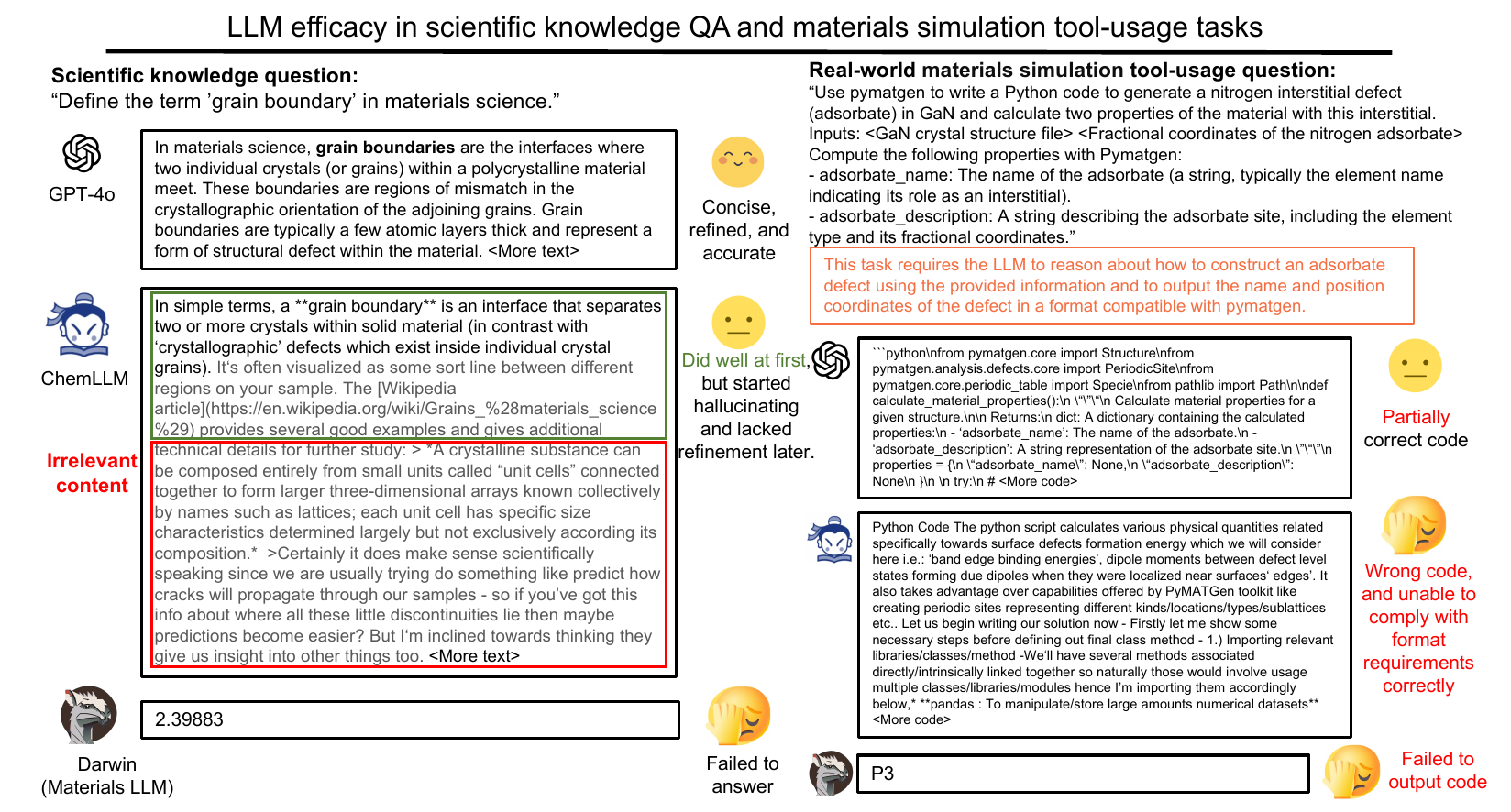}
	\caption{Performance of large language models on scientific knowledge and materials simulation tasks. Representative responses of three LLMs to (a) scientific knowledge questions and (b) code generation tasks for materials simulation tools.
}\label{fig:1}
\end{figure*}

Recent developments introduced LLM-based agents~\citep{atomagents, chemcrow, D4DD00252K} that interface with existing software tools and collaborate with researchers to address complex scientific tasks. While effective for certain applications, these approaches remain constrained by their dependence on human-written code, limiting their ability to autonomously solve scientific problems. To assess current capabilities, we benchmarked LLMs on two representative scenarios in materials science: knowledge-based question answering and tool-driven code generation (Fig.~\ref{fig:1}). In the knowledge QA task, GPT-4o~\citep{gpt4o} produced concise, accurate, and well-structured responses, whereas ChemLLM~\citep{chemllm}, a domain-specific model for materials chemistry, generated lengthy and error-prone outputs. Performance degraded further in the code generation task: GPT-4o failed to produce functional implementations, and domain-specific LLMs were unable to generate executable Python code for materials simulations. These results highlight a critical gap; domain-specialized models do not yet surpass general-purpose LLMs, and neither class of models meets the robustness and autonomy required for scientific computation.

To address this gap, we first develop benchmarks to evaluate the capability of LLMs to use materials science tools. Existing scientific LLM benchmarks have largely emphasized reading comprehension~\citep{MaScQA,SciEval, SciAssess, SciQAG} and materials property prediction~\citep{rubungo2025, D4SC04401K, Li2025-ed}, but have overlooked the critical ability to generate code for performing physically meaningful calculations. For example, benchmarks such as MaScQA~\citep{MaScQA}, SciEval~\citep{SciEval}, SciAssess~\citep{SciAssess}, SciQAG~\citep{SciQAG}, and SciFIBench~\citep{roberts2024scifibench} primarily assess knowledge understanding through QA or multiple-choice formats, focusing on literature or image interpretation in uni-modal or multi-modal contexts. These valuable benchmarks are typically general-purpose, with only limited coverage of materials science tasks. Additional benchmarks target property prediction, including LLM4Mat-Bench~\citep{rubungo2025}, MatText~\citep{mattext}, and recent efforts by Joren et al.~\citep{D4SC04401K} and Li et al.~\citep{Li2025-ed}. In contrast, our work focuses on applications that integrate LLMs with physically based simulation tools, and we introduce a benchmark explicitly designed to assess both the autonomous use of materials science software and the performance of combined LLM–tool systems. This design establishes a distinct and necessary benchmark direction for advancing AI-driven scientific discovery.

Recent efforts have introduced benchmarks for scientific code generation, yet their coverage of materials science remains limited. SciCode~\citep{tian2024scicode} established a benchmark for code generation across multiple scientific disciplines, encompassing mathematics, physics, chemistry, biology, and materials science. This benchmark consists of 80 primary problems and 338 sub-problems, of which only 16\% pertain to materials science. HoneyComb~\citep{honeycomb} reframed scientific QA pairs into code completion tasks (similar to Toolformer~\citep{toolformer}) and demonstrated improved LLM performance on the MaScQA dataset. This approach does not assess real-world tool-usage capabilities (the focus of our work). More recently, Shi et al.~\citep{Shi2025gk} proposed the LAMMPS-Expert Question with Score (LEQS) dataset, which evaluates LLM-generated scripts for the widely-used molecular dynamics package LAMMPS~\citep{LAMMPS}. However, this benchmark is constrained to predefined metrics (e.g., grammatical, logical, and parameter errors) and single-task evaluations based on curated datasets. In contrast, our work introduces a benchmark specifically designed to assess LLMs’ ability to operate materials science tools through a multi-level reasoning and testing framework, thereby providing a more comprehensive evaluation of their capacity for autonomous scientific computation.

The construction of high-quality question–answer (QA) pairs in specialized domains such as biology, mathematics, and materials science has traditionally been a labor-intensive, time-consuming, and costly process. 
The development of LLMs has enabled the automatic generation of high-quality QA pairs based on factual , open-ended , and multiple-choice questions~\citep{maity}. 
While LLMs can support QA generation across scientific fields, the knowledge focus varies by domain; (i) biomedical QAs emphasizes professional literature, terminology, genomic data, and clinical reasoning~\citep{medical1, medical2}, (ii) chemistry emphasizes reaction mechanisms, experimental workflows, and molecular structures~\citep{chemistry1,chemistry2}, (iii) materials science emphasizes composition–structure–property relationships~\citep{rubungo2025, mattext}, and (iv) mathematics emphasizes logical reasoning~\citep{math1, math2}. 
A common strategy in QA benchmark development is the use LLMs to generate structured instructions, followed by fine-tuning smaller models to scale up QA generation; examples include Darwin in  materials science~\citep{materials1}, m-KAILIN in medicine~\citep{mediacal_qa}, and SCQA in chemistry~\citep{D4DD00307A}.

In contrast, our work targets automatic QA generation for materials science tools, with a focus on code repositories rather than unstructured scientific knowledge. 
To this end, we designed question templates inspired by HumanEval~\citep{humaneval} and BigCodeBench~\citep{bigcodebench} to construct \texttt{pymatgen\_code\_qa}, in order to evaluate an LLMs’ ability to generate code for the widely used, open-source materials science Python library \texttt{pymatgen} library (for data processing, analysis, and simulation). 
We further developed \texttt{pymatgen\_doc\_qa}, a benchmark based on questions about \texttt{pymatgen}’s documentation, to test whether LLMs understand the scientific principles underlying computational materials codes.

Building on these components, we introduce MatTools, a comprehensive multi-step reasoning benchmark designed to evaluate LLM capabilities in materials science tool utilization. MatTools integrates two complementary elements: (1) a large-scale simulation tool QA benchmark with 69,225 QA pairs derived from the \texttt{pymatgen}~\citep{ONG2013314} codebase and documentation, and (2) a real-world tool-usage benchmark comprising 49 questions (138 tasks) requiring the generation of functional Python code for defect property calculations, constructed from property test files in the \texttt{pymatgen-analysis-defects}~\citep{pymatgen-defects} library. MatTools addresses key limitations of existing benchmarks through three design principles: (1) Automated data synthesis: real-world tool-usage benchmarks are generated directly from property test files, eliminating the need for manual data collection or expert annotation. (2) Comprehensive dual-benchmark design: by combining QA and real-world tool-usage benchmarks, MatTools enables systematic evaluation of both knowledge comprehension and practical tool-use capabilities, for both standalone LLMs and retrieval-augmented generation (RAG) agent systems. (3) Secure and standardized evaluation: Docker~\citep{docker} sandboxing ensures safe and reproducible execution of LLM-generated code.

Using this framework, we conduct multi-level evaluations of LLM performance in materials science tool utilization. Our experiments yield three key insights: (1) Generalists outshine specialists: General-purpose LLMs (such as GPT-4o and Qwen2.5 series~\citep{qwen25}) significantly outperform domain-specific materials science LLMs in  knowledge QA tasks (80\% vs. $<$32\% accuracy for general-purpose vs. domain-specific LLMs). (2) AI knows AI: incorporating LLM-generated documentation as a retrieval source in RAG systems markedly improves code generation, increasing the success rate of runnable code by 47.8\% and overall task success by 115.7\% relative to GPT-4o alone. (3) Simpler is better: a self-reflection LLM-doc RAG agent system, using only LLM-generated documentation and multi-round reflection, surpasses more complex methods such as agentic RAG (with task decomposition, NER, and reranking) and the state-of-the-art GraphRAG approach LightRAG~\citep{lightrag}. Our method achieves 58.8\% and 149\% higher task success rates compared to agentic RAG and LightRAG, respectively. Even a single LLM+RAG system outperforms them by 13.7\% and 78.3\%. These findings highlight both the current limitations of domain-specific LLMs and the potential of leveraging LLM-generated documentation and self-reflection mechanisms to enhance tool-use capabilities in materials science.

\begin{figure*}[htbp]
	\centering
	\includegraphics[width=\textwidth]{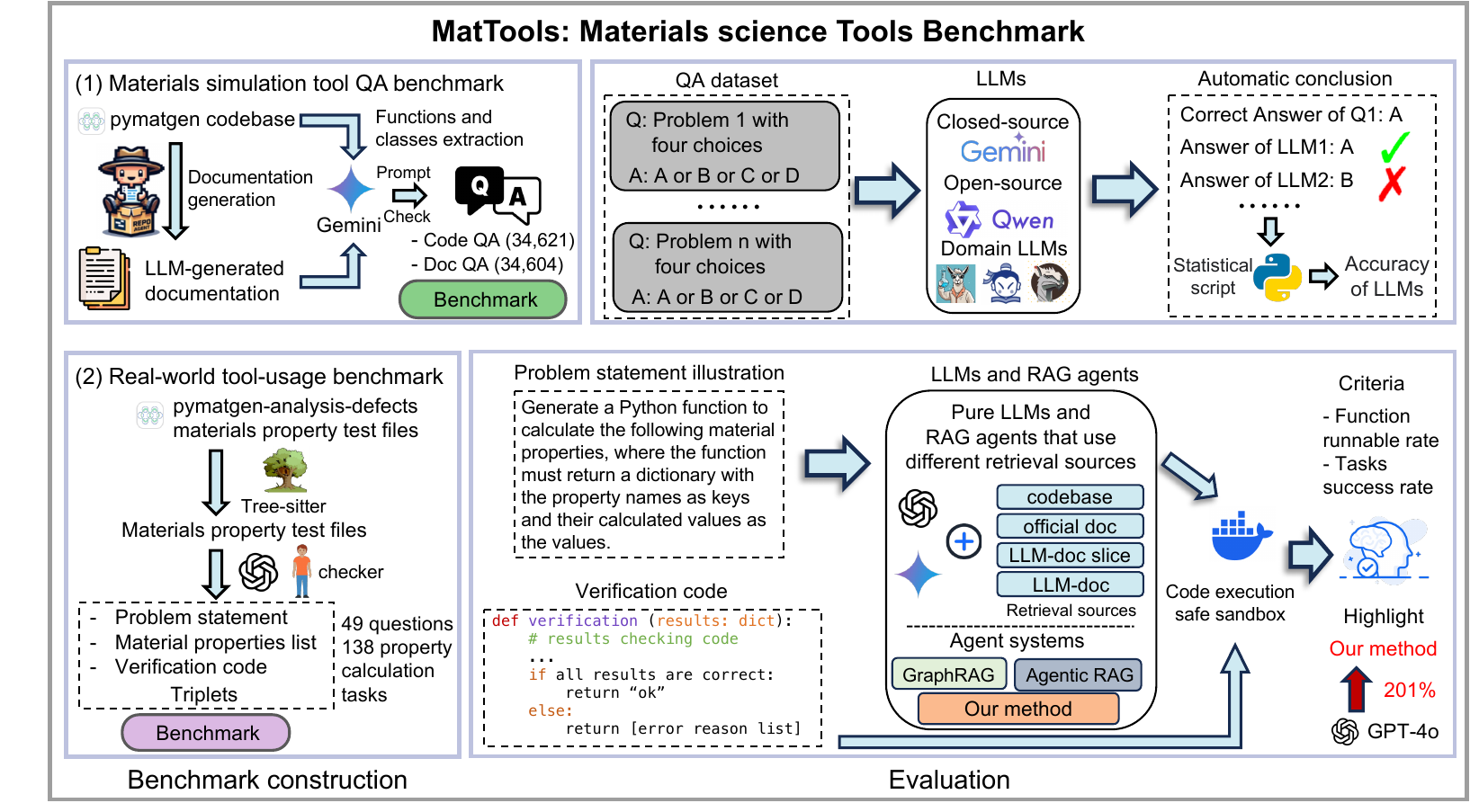}
	\caption{Overview of MatTools benchmarking framework. The upper panel illustrates the pipeline for constructing and applying the QA benchmark: \texttt{RepoAgent} extracts code snippets from the \texttt{pymatgen} library, and Gemini-2.0-flash generates documentation. Using these inputs, Gemini-2.0-flash produces the \texttt{pymatgen\_code\_qa} and \texttt{pymatgen\_doc\_qa} benchmarks, in which LLMs are evaluated by answering multiple-choice questions (A–D). The lower panel depicts the real-world tool-usage benchmark: \texttt{Tree-sitter}~\citep{treesitter} extracts test functions from materials property test files, and GPT-4o generates question–property–validation triples. Code generated by LLM-based systems is then executed and validated within a secure Docker sandbox.}\label{fig:2}
\end{figure*}

\begin{figure*}[htbp]
	\centering
	\includegraphics[width=\textwidth]{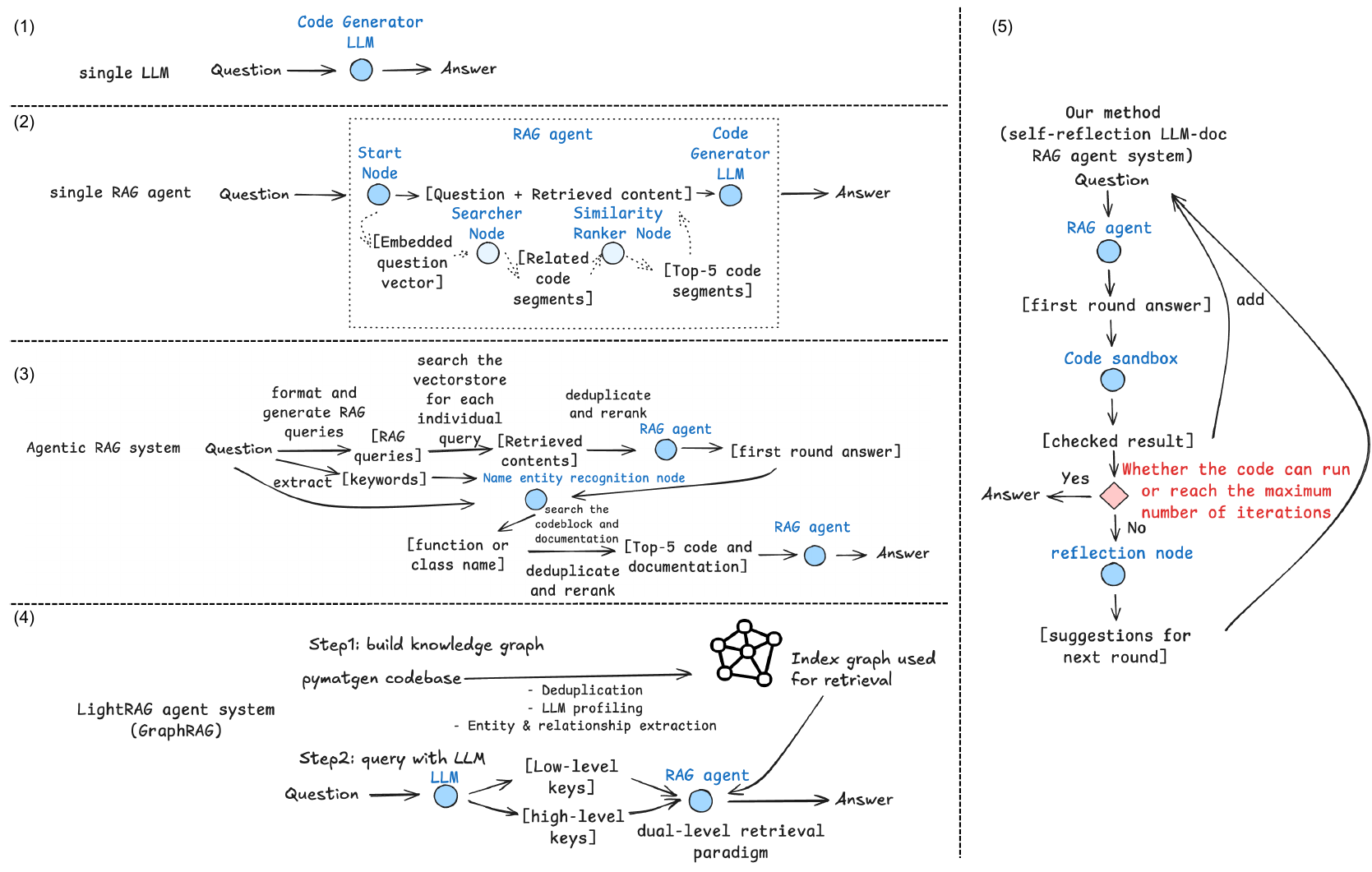}
	\caption{Illustration of the five benchmarked LLM-based systems for real-world tool-usage.}\label{fig:3}
\end{figure*}

\section{MatTools}
MatTools comprises two complementary components: (i) a materials simulation tool QA benchmark (Section~\ref{sec:qa_benchmark}), which assesses LLMs’ knowledge and comprehension of materials science concepts, and (ii) a real-world tool-usage benchmark (Section~\ref{sec:real_world_benchmark}), which evaluates their ability to generate functional code for scientific computation. For each benchmark, we describe the methodology for data collection and synthesis, as well as the testing frameworks developed to systematically assess LLM performance in tool usage (Fig.~\ref{fig:2}).

\subsection{\textbf{Materials simulation tool QA benchmark}}\label{sec:qa_benchmark}
We selected \texttt{pymatgen} as the primary data source for constructing the QA benchmark. The repository was processed using \texttt{Repoagent}~\citep{luo-etal-2024-repoagent} in four steps: (1) \textbf{Repository parsing}: the codebase was analyzed to construct a hierarchical project tree with the repository as the root node and directories/Python files as intermediate nodes; (2) \textbf{Structure extraction}: classes and functions were identified as leaf nodes, and caller-callee relationships were recorded to form a directed acyclic graph (DAG); (3) \textbf{Documentation generation}: \texttt{Gemini-2.0-flash}~\citep{gemini2series}, guided by specialized texttt{RepoAgent} prompts (Supplementary Material~\ref{sec:a1.1}), generated documentation for each code segment; and (4) \textbf{Dataset creation}: two datasets, \texttt{pymatgen\_code} and \texttt{pymatgen\_doc}, were compiled, each containing 7,192 datapoints derived from code segments and their corresponding documentation (See Methods Section for more details).

We designed two prompt templates (Supplementary Material~\ref{sec:a1.2}) to generate QA pairs from the \texttt{pymatgen\_code} and \texttt{pymatgen\_doc} datasets. \texttt{Gemini-2.0-flash} produced up to 5 distinct questions per datapoint (code segment or documentation), with fewer questions generated when the content was insufficient to support 5 meaningful items. Each QA pair consisted of a question with four answer options (A–D), requiring LLMs to respond with only A, B, C, or D. This process yielded two benchmarks: \texttt{pymatgen\_code\_qa} with \textbf{34,621} QA pairs and \texttt{pymatgen\_doc\_qa} with \textbf{34,604} QA pairs (see Supplementary Material~\ref{sec:a1.3}). 
While \texttt{pymatgen\_code\_qa} primarily evaluates understanding of the physical basis of various descriptors in the pymatgen code (e.g., variables corresponding to specific physical quantities), \texttt{pymatgen\_doc\_qa} reflects deeper conceptual understanding by testing whether a LLM grasps the physical meaning of variables as described in the documentation.

To systematically evaluate LLM comprehension of materials simulation tools and the effect of model scale, we benchmarked 9 general-purpose LLMs, including 3 widely used closed-source models and 6 Qwen2.5 open-source models spanning different parameter sizes. In addition, we evaluated 3 domain-specific LLMs trained for materials chemistry (Supplementary Section~\ref{sec:a1.4}), which have shown strong performance in literature understanding and property prediction, to determine whether such specialization extends to tool knowledge and instruction following. Model performance was quantified by accuracy, defined as the proportion of correctly answered questions, enabling direct comparison of comprehension across general-purpose and domain-specific LLMs.

\subsection{\textbf{Real-world tool-usage benchmark}}\label{sec:real_world_benchmark}
The real-world tool-usage benchmark is designed to closely replicate the workflow of a computational materials scientist, thereby testing both the practical skills and conceptual understanding required to use sophisticated software tools. 
Our benchmark is not intended to ``replace'' a materials scientist but rather provides a framework for automating the intermediate engineering steps that arise during the scientific process. 
In this way, the benchmark reflects the actual questions a materials scientist would ask during problem-solving, rather than abstract, high-level research questions or application-driven queries. 
The benchmark emulates the scientific workflow through three key stages: 
(1) Problem Formulation: a scientist begins with a research objective (e.g., ``Determine the energy of an octahedral self-interstitial point defect in GaN'') - the benchmark then provides the LLM with a problem statement that encapsulates this scientific goal. 
(2) Code Implementation: the scientist typically writes a script using a specialized library (\texttt{pymatgen} here) to perform the required calculation - the benchmark then evaluates the ability of the LLM to generate such functional Python code as a direct test of tool-usage ability. 
(3) Execution and Verification: our benchmark automatically executes the code in a secure Docker sandbox, and assesses the outcomes according to two criteria: (i) whether the code runs successfully (Function Runnable Rate), and (ii) whether the result is scientifically correct (Task Success Rate).

Success in this benchmark requires more than generic code generation. 
The LLM must demonstrate a working knowledge of materials science concepts as implemented in \texttt{pymatgen}, including: 
(1) Crystallography: correct construction and use of \texttt{Structure} objects to represent crystal lattices. 
(2) Defect Physics: proper differentiation between defect types (e.g., interstitials, vacancies) and the corresponding functions to model them. 
(3) Tool-Specific API Knowledge: accurate Navigation of the \texttt{pymatgen} library to select and apply the appropriate classes and methods. 
The difficulty of this benchmark is underscored by the performance of GPT-4o, which achieves a Task Success Rate of only 18.36\%. 
Overall, the benchmark assesses the practical ability of LLMs to translate a scientific problem into a valid computational solution, reflecting the essence of ``real-world'' tool usage in computational materials science. 

We developed an automated pipeline using LLMs to transform materials property test code into triplets, enabling the generation of a real-world tool-usage benchmark with \textbf{49 questions} (\textbf{138 tasks}), each corresponding to a specific material property calculation. This automated approach ensures scalability, reproducibility, and scientific relevance. Further methodological details of the benchmark construction are provided in the Methods section. A testing framework was then developed to integrate the synthesized benchmark data with the Docker sandbox to systematically evaluate 5 different LLM-based approaches. For each evaluation, the framework provides the problem statement from the generated triplets as input to an LLM-based system, which attempts to produce Python code for calculating the target material properties. The generated code is executed within the Docker sandbox to produce a material property dictionary. This output is then validated by executing the verification script in the sandbox, which checks the correctness of the results. We tested both direct LLMs and more advanced agentic systems for complex code generation tasks to address the challenges of real-world materials simulation tool usage. We designed and evaluated five distinct systems (Fig.~\ref{fig:3} and Supplementary Section~\ref{sec:a2.2}): (1) a single LLM, (2) a single RAG agent using \texttt{pymatgen} source code or documentation retrieval, (3) a multi-agent RAG system incorporating task decomposition, NER, and re-ranking, (4) a GraphRAG system leveraging structured knowledge representations (implemented using the state-of-the-art LightRAG method), and (5) our self-reflection LLM-doc RAG system, which combines LLM-generated documentation retrieval with iterative refinement. Performance was quantified by the number of runnable functions (49 in total) and successful tasks (138 in total), as determined by verification of the generated code within the Docker sandbox for each LLM-based system.

\section{Materials simulation tool QA benchmark results}
\subsection{Quantitative results}
\paragraph*{\textbf{Results on non-reasoning LLMs}}
Table~\ref{table:1} summarizes the performance of different LLMs on the materials simulation tool QA benchmarks. General-purpose LLMs--both closed-source and open-source--consistently outperform domain-specific materials chemistry models in understanding and reasoning about simulation tool knowledge. Leading general models, such as Gemini-1.5-Pro, Qwen2.5-32B-Instruct, and Qwen2.5-72B-Instruct, achieve over 80\% accuracy on both code- and documentation-based QA tasks. In contrast, specialized materials chemistry models (ChemDFM-v1.5-8B, ChemLLM-7B-Chat-1\_5-DPO, and Darwin-1.5-7B) perform substantially worse, with accuracies $\sim$30\%, and in one case close to zero. The poor performance of ChemLLM-7B-Chat-1\_5-DPO and Darwin-1.5-7B is largely attributable to limited instruction-following ability, often leading to improperly formatted answers (e.g., ``$\langle answer \rangle$Option$\langle /answer \rangle$''). Overall, general-purpose LLMs demonstrate superior instruction adherence, broader knowledge coverage, and stronger generalization for materials simulation tools compared with domain-specific alternatives. Among open-source models, performance improves systematically with increasing model size, as illustrated by the Qwen2.5 family.

\paragraph*{\textbf{Results on reasoning LLMs}}
We further evaluated several reasoning LLMs on the QA benchmarks (see Supplementary Table~\ref{tab:reasoning_qa_results}). 
Since reasoning models generate intermediate ``thought'' processes before producing an answer, we applied ``regular expressions'' to extract the final choice. 
The reasoning model qwq-32B achieved 84.27\% accuracy on the \texttt{pymatgen\_code\_qa} benchmark, surpassing the best-performing non-reasoning model, Qwen2.5-72B-Instruct (81.36\%). 
This indicates that the fixed-choice QA format does not hinder reasoning models and may even highlight their strengths. 
Additional tests, on smaller Qwen3-series reasoning models, show that accuracy scales positively with model size, consistent with trends observed in general-purpose LLMs.

\begin{table}[htbp]
	\centering
	\caption{Accuracy of non-reasoning LLMs on the materials simulation tool QA benchmarks, evaluated on code-based (\texttt{pymatgen\_code\_qa}) and documentation-based (\texttt{pymatgen\_doc\_qa}) tasks.}
	\begin{tabular}{c cc}
		\toprule
		\multirow{2}{*}{\textbf{Model}} & \multicolumn{2}{c}{\textbf{Accuracy (\%)}} \\
		\cmidrule(lr){2-3}
		& \textbf{Code} & \textbf{Doc} \\
		\midrule
		\multicolumn{1}{c}{\textbf{Closed-source LLMs}} \\
		\midrule
		Gemini-1.5-Flash & 75.59 & 76.37 \\
		Gemini-1.5-Pro & \textbf{80.60} & \textbf{82.90} \\
		Gemini-2.0-Flash & 73.85 & 82.76 \\
		\midrule
		\multicolumn{1}{c}{\textbf{Open-source LLMs}} \\
		\midrule
		Qwen2.5-7B-Instruct & 74.37 & 75.79 \\
		Qwen2.5-14B-Instruct & 76.72 & 80.26 \\
		Qwen2.5-Coder-14B-Instruct & 77.44 & 79.44 \\
		Qwen2.5-32B-Instruct & 80.03 & \textbf{82.01} \\
		Qwen2.5-Coder-32B-Instruct & 79.18 & 79.85 \\
		Qwen2.5-72B-Instruct & \textbf{81.36} & 81.82 \\
		\midrule
		\multicolumn{1}{c}{\textbf{Materials chemistry LLMs}} \\
		\midrule
		ChemDFM-v1.5-8B & \textbf{32.35} & \textbf{30.20} \\
		ChemLLM-7B-Chat-1\_5-DPO & 0.18 & 0.13 \\
		Darwin 1.5-7B & 0.01 & 0.01 \\
		\bottomrule
	\end{tabular}\label{table:1}
\end{table}

\subsection{Qualitative analysis}
\paragraph*{\textbf{Impact of instruction-following ability on LLM performance}}
As shown in Table~\ref{table:1}, general-purpose LLMs exhibit stronger instruction-following abilities in materials simulation tools compared with domain-specific LLMs. 
A key reason for the poor performance of domain-specific LLMs is their inability to follow the required output format (Supplementary Figs.~\ref{fig:qa_results1} and~\ref{fig:qa_results2}). 
Training objectives - general models are usually trained for instruction following (e.g., the Qwen Instruct models we tested underwent systematic instruction tuning and alignment to enhance their ability to follow diverse prompts, generate structured outputs (e.g., JSON), and handle long contexts ($>$8k tokens)). In contrast, domain-specific LLMs are primarily tuned for materials science tasks but lack broad instruction-following optimization. Training/fine-tuning data - materials science models emphasize specialized content such as terminology, property prediction, or literature-based knowledge extraction; their training emphasizes domain knowledge rather than tool usage. While the three domain models did train on materials data, they did not train on any materials science coding datasets. In contrast, models like Qwen are exposed to broader data distributions, including code and mathematics, which directly benefits tool-usage tasks. Generalization ability - while domain-specific models often suffer from weaker transferability (struggling when faced with queries outside or slightly beyond their training distribution (Fig.~\ref{fig:1})), general-purpose LLMs, trained on larger and more diverse corpora, demonstrate stronger robustness and adaptability to varied question formats.

\begin{figure}[htbp]
	\centering
	\includegraphics[width=0.45\textwidth]{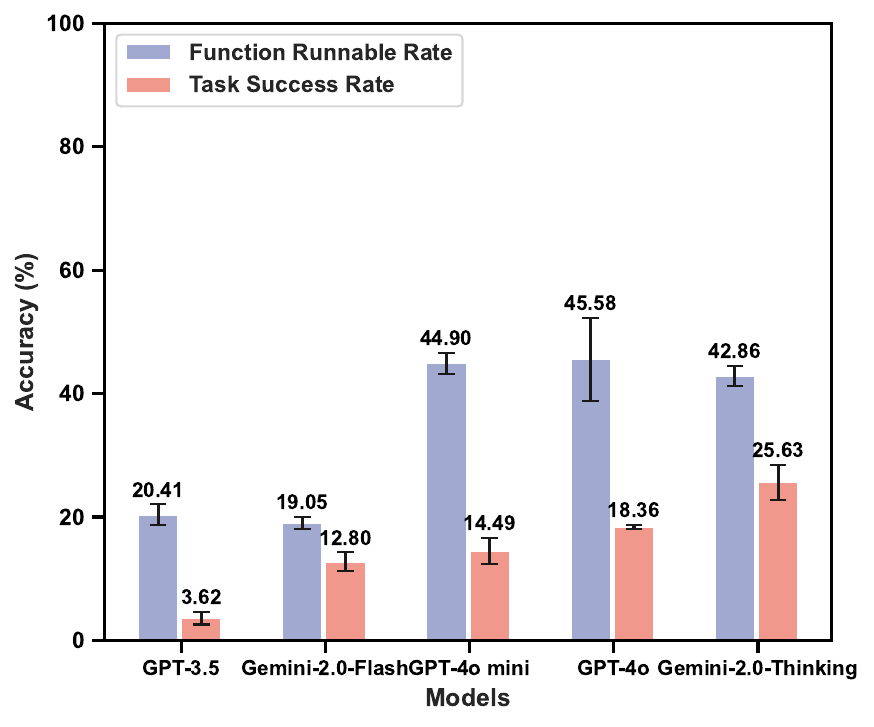}
	\caption{Performance comparison of different LLMs on the real-world tool-usage benchmark. Reported values represent mean performance across three independent trials, with error bars indicating the corresponding standard deviations.}\label{fig:4}
\end{figure}

\begin{figure*}[htbp]
	\centering
	\includegraphics[width=0.85\textwidth]{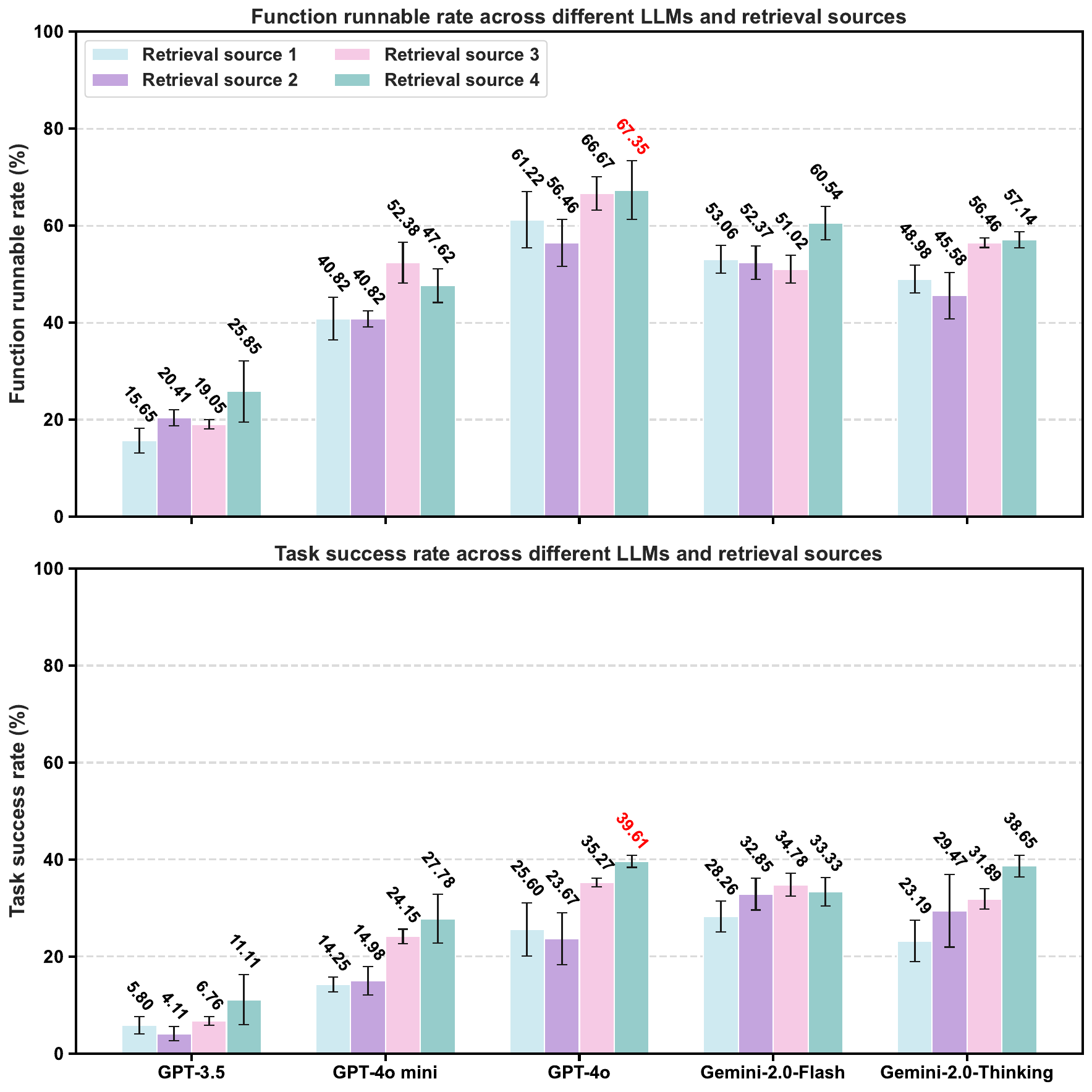}
	\caption{Comparative performance of a single RAG agent using different LLMs and retrieval sources on the real-world tool-usage benchmark. Retrieval sources include: (1) the \texttt{pymatgen} codebase, (2) the official \texttt{pymatgen} documentation (character-level split), (3) LLM-generated documentation (semantically segmented), and (4) LLM-generated documentation split by function and class. Error bars indicate standard deviation across three independent experimental runs; displayed values represent mean performance metrics from these trials.}\label{fig:5}
\end{figure*}

\begin{figure}[htbp]
	\centering
	\includegraphics[width=0.45\textwidth]{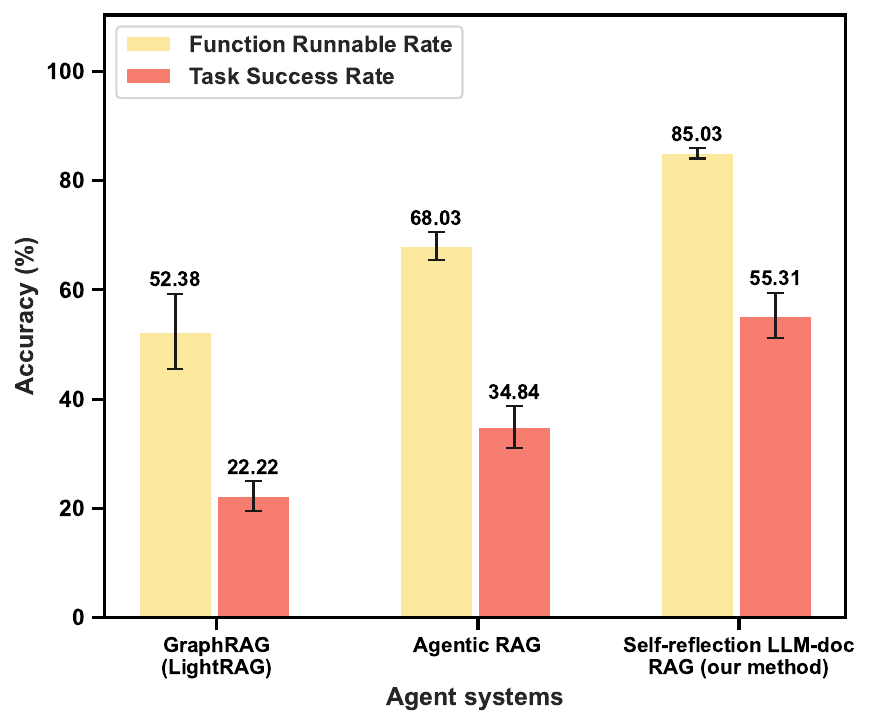}
	\caption{Comparative performance of advanced RAG agent systems on the real-world tool-usage benchmark. All systems used GPT-4o as the base model for code generation.
	}\label{fig:6}
\end{figure}

\paragraph*{\textbf{Impact of fine-tuning on LLM performance}}
We further examined the role of fine-tuning, since domain-specific LLMs are typically derived from general-purpose models (e.g., ChemDFM from Llama2-13B, ChemLLM from InternLM2-7B, and Darwin from Llama3-8B). 
Interestingly, fine-tuning can degrade performance. 
ChemLLM and Darwin, for example, frequently produced extraneous or nonsensical tokens when faced with unfamiliar queries (Fig.~\ref{fig:1}); this behavior is rarely observed in their base models. 
For example, when asked to define ``grain boundary'' in Fig.~\ref{fig:1},  Llama3-8B responded correctly with a standard scientific definition, whereas Darwin produced the meaningless response ``2.39883''. 
This indicates that fine-tuning, while improving domain-specific knowledge (e.g., property prediction, crystal structure identification, molecular recognition), can compromise broader abilities (including coding). 
Our previous work~\citep{liu2024llms, LIU2024240} similarly observed that small, fine-tuned LLMs often overfit formatting patterns in their fine-tuning data. 
These findings are consistent with a recent study~\citep{MSQA}, which developed MSQA, a benchmark of 1,757 graduate-level materials science questions, and reported that domain-specific LLMs systematically underperform due to overfitting and distributional shifts. 
This impacts the capabilities of models for our benchmark which focuses on the ability to use materials science computational tools; an important area, largely overlooked in the literature.

These results underscore the superior robustness, instruction-following ability, and generalization of general-purpose LLMs compared to domain-specific models in materials simulation tool QA tasks. Based on this observation, we focus the remainder of our analysis on general-purpose LLMs in the real-world tool-usage benchmark.

\section{Real-world tool-usage benchmark results}
To assess LLM performance on the real-world tool-usage benchmark, we designed three types of tests. 
First, we queried LLMs directly with benchmark questions. In this setting, both the Function Runnable Rate and Task Success Rate were low ($<$50\%), highlighting the difficulty of generating fully functional, scientifically correct code without additional support. Next, we tested if RAG could improve performance. We evaluated four different retrieval sources (lower panel of Fig.~\ref{fig:2}) and found that using LLM-generated documentation as the RAG retrieval source yielded the best results. Based on this observation, we designed a simple self-reflection RAG agent system. This system iteratively refines code by analyzing the output of each execution round and generating subsequent rounds of code based on reflection (see Section \ref{sec:real_world_benchmark}). The approach achieved substantial improvements: a 149\% increase in Task Success Rate over the state-of-the-art GraphRAG method (LightRAG), a 58.8\% improvement over the agentic RAG system with task decomposition, NER, and reranking, and a 201\% improvement over GPT-4o alone (see Supplementary Material~\ref{sec:a2.3} for additional details and examples). 
Finally, analysis of failure modes indicates that our self-reflection RAG agent effectively handles the ``how'' of code generation (syntax, imports, and API calls) but can still struggle with the ``what'' (i.e., the underlying scientific and algorithmic logic). 
In contrast, other LLMs and agent systems frequently fail even at the ``how'' stage, demonstrating the advantage of iterative reflection combined with RAG.

\subsection{Results of testing single LLM system}
Fig.~\ref{fig:4} presents the performance of different LLMs on the real-world tool-usage benchmark. GPT-3.5 achieves a Function Runnable Rate of only 20.41\% and a Task Success Rate of 3.62\%. Even the top-performing model, GPT-4o, reaches only 45.58\% Function Runnable Rate and 18.36\% Task Success Rate. Among reasoning models, Gemini-2.0-flash-thinking-exp-01-21 attains the highest Task Success Rate (25.63\%) but achieves a Function Runnable Rate of just 42.86\%. All tested models exhibit low Function Runnable Rates and Task Success Rates, highlighting that current mainstream LLMs, including reasoning-enabled models, struggle to perform complex materials science tool usage tasks reliably. The low Function Runnable Rates indicate that generated code often requires modification to execute successfully, while the low Task Success Rates show that even runnable code frequently produces incorrect or unreliable results. To address these limitations, we next evaluated the effectiveness of RAG for enhancing LLM performance in materials science tool usage.

\subsection{Results of testing a single RAG agent with different retrieval sources}
Fig.~\ref{fig:5} compares the performance of a single RAG agent using various LLMs and retrieval sources on the real-world tool-usage benchmark. Among the four retrieval sources, using LLM-generated documentation split based on function and class yielded the best results. Specifically, GPT-4o with this retrieval source achieved a Function Runnable Rate of 67.35\% and a Task Success Rate of 39.61\%, representing improvements of 47.8\% and 115.7\%, respectively, over GPT-4o alone, and 19.3\% and 67.3\% over GPT-4o with the official documentation. 
Comparing official versus LLM-generated documentation reveals that the latter provides more detailed, structured information. 
The LLM-generated documents describe the purpose and usage examples for all APIs under each class, organized in a markdown format similar to AI-generated repository documentation tools such as DeepWiki.
 This structure lowers the barrier for understanding and includes more practical examples of API usage which, we speculate, is a key factor driving  improved RAG performance. 
These results indicate that LLM-generated documentation enhances content retrieval quality and substantially improves real-world tool-usage outcomes.

\subsection{Results of testing advanced RAG agents}

Building on these results, we designed a self-reflection LLM-doc RAG agent system that uses LLM-generated documentation segmented by function and class as the retrieval source and incorporates iterative reflection to provide feedback on generated code. Fig.~\ref{fig:6} compares the performance of this system with other mainstream RAG agent approaches on the real-world tool-usage benchmark, using GPT-4o as the base LLM for all systems. The self-reflection LLM-doc RAG agent achieved a 26.3\% improvement in Function Runnable Rate and a 39.6\% improvement in Task Success Rate compared to a RAG system without self-reflection. Notably, the agentic RAG system with task decomposition, NER, and reranking attained a lower Task Success Rate than GPT-4o combined with LLM-doc RAG, while the GraphRAG method (LightRAG) performed even worse. These results indicate that combining LLM-generated documentation with iterative self-reflection outperforms mainstream RAG approaches in materials science tool usage tasks, even though methods like LightRAG and agentic RAG typically excel in other domains. Compared with using GPT-4o alone, the self-reflection LLM-doc RAG system achieves substantial gains, improving Function Runnable Rate by 86.6\% and Task Success Rate by 201.3\%.

\begin{table*}[htbp]
    \caption{Distribution of error types across different LLMs and agent-based methods on the real-world tool-usage benchmark.}\label{tab:error_breakdown}
    \centering
    \resizebox{\textwidth}{!}{
    \begin{tabular}{cccccccc}
    \toprule
    \textbf{Category} & \textbf{Model or Method} & \textbf{Import Error} & \textbf{API Hallucination} & \textbf{Incorrect Parameter} & \textbf{Syntax Error} & \textbf{Type Mismatch} & \textbf{Logic Error} \\
    \midrule
    \multirow{5}{*}{Pure LLM} 
    & gemini-2.0-flash & 14 & 4 & 3 & 17 & 4 & 3 \\
    & gemini-2.0-flash-thinking & 18 & 8 & 0 & 1 & 5 & 8 \\
    & gpt-3.5 & 10 & 1 & 5 & 20 & 5 & 6 \\
    & gpt-4o & 16 & 7 & 2 & 1 & 7 & 5 \\
    & gpt-4o-mini & 14 & 7 & 4 & 2 & 5 & 8 \\
    \midrule
    \multirow{12}{*}{RAG}
    & \makecell[c]{gemini-2.0-flash-thinking\\(retrieval source 1)} & 12 & 6 & 3 & 3 & 7 & 7 \\
    & \makecell[c]{gemini-2.0-flash-thinking\\(retrieval source 2)} & 15 & 6 & 4 & 4 & 3 & 7 \\
    & \makecell[c]{gemini-2.0-flash-thinking\\(retrieval source 3)} & 12 & 3 & 3 & 3 & 5 & 7 \\
    & \makecell[c]{gemini-2.0-flash-thinking\\(retrieval source 4)} & 10 & 5 & 5 & 1 & 4 & 5 \\
    & \makecell[c]{gpt-4o\\(retrieval source 1)} & 9 & 5 & 0 & 2 & 9 & 9 \\
    & \makecell[c]{gpt-4o\\(retrieval source 2)} & 10 & 5 & 0 & 1 & 10 & 9 \\
    & \makecell[c]{gpt-4o\\(retrieval source 3)} & 7 & 2 & 4 & 3 & 5 & 9 \\
    & \makecell[c]{gpt-4o\\(retrieval source 4)} & 9 & 2 & 3 & 6 & 2 & 8 \\
    & \makecell[c]{gpt-3.5\\(retrieval source 1)} & 10 & 6 & 8 & 12 & 5 & 2 \\
    & \makecell[c]{gpt-3.5\\(retrieval source 2)} & 13 & 2 & 13 & 11 & 3 & 5 \\
    & \makecell[c]{gpt-3.5\\(retrieval source 3)} & 11 & 2 & 7 & 15 & 5 & 3 \\
    & \makecell[c]{gpt-3.5\\(retrieval source 4)} & 12 & 4 & 1 & 20 & 4 & 2 \\
    \midrule
    \multirow{3}{*}{Agent system}
    & LightRAG & 17 & 6 & 4 & 1 & 4 & 3 \\
    & Agentic RAG & 9 & 3 & 2 & 2 & 7 & 11 \\
    & Our method & 5 & 1 & 0 & 2 & 5 & 12 \\
    \bottomrule
    \end{tabular}}
\end{table*}

\subsection{Analysis of model failure modes}
In our testing workflow, error logs were automatically recorded, allowing systematic analysis of failure cases. 
We sampled 748 errors across 20 LLMs and agent-based methods over 49 tasks and manually categorized them into six types: (1) Import Error (missing necessary imports), (2) API Hallucination (calling non-existent functions, classes, or methods in \texttt{pymatgen}, (3) Incorrect Parameter (providing invalid parameters to otherwise valid APIs), (4) Syntax Error (Python code that is syntactically invalid and fails to execute), (5) Type Mismatch (execution returns an object of an unexpected type relative to verification requirements), and (6) Logic Error (code that is syntactically correct and API-valid but fails to produce correct results due to flawed algorithmic reasoning).

Table~\ref{tab:error_breakdown} summarizes the distribution of these error types. Pure LLM Less advanced models (e.g., GPT-3.5, Gemini-2.0-flash) frequently fail due to Syntax Errors, indicating difficulty in producing executable Python code that meets verification requirements. More advanced models (e.g., GPT-4o, Gemini-2.0-flash-thinking) exhibit far fewer syntax errors, but their failures shift toward Import Errors and Logic Errors. 
This suggests progress in surface-level code generation, yet persistent challenges in contextual reasoning and task-specific implementation. 
Incorporating RAG substantially reduces API Hallucination by grounding models in documentation. For example, GPT-4o with LLM-generated documentation reduced API hallucinations from 7 to 2. 
However, RAG does not resolve all error types: GPT-3.5 continues to fail predominantly with Syntax Errors across retrieval sources, while GPT-4o still exhibits Logic Errors and Type Mismatches, even when retrieval is provided. 
Agent-based approaches show progressive improvement. 
LightRAG, despite leveraging structured knowledge graphs, struggled with Import Errors (17 cases), suggesting insufficient coverage of dependency information. 
Agentic RAG  reduced such foundational issues but introduced a higher incidence of Logic Errors (11 cases), reflecting the added complexity of multi-agent coordination. 
Our self-reflection LLM-doc RAG agent achieved the most robust performance, with the lowest counts of Import Errors (5), API Hallucinations (1), and Incorrect Parameters (0). 
This demonstrates the effectiveness of iterative self-reflection in correcting surface-level and structural mistakes. 
However, its errors were concentrated in the Logic Error category (12 cases), underscoring a key limitation: while the system effectively addresses the ``how'' of code generation (syntax, imports, API calls), it continues to struggle with the ``what'' (scientific reasoning and algorithmic correctness). 
This  highlights a critical direction for future work, enhancing the logical reasoning capabilities of LLM-based agents for scientific tool usage.

\section{Conclusion}
We introduced MatTools, a dual-level multi-step reasoning benchmark consisting of a QA benchmark and a real-world tool-usage benchmark, designed to evaluate and advance the capabilities of LLMs in computational materials science. 
The real-world benchmark was validated through expert review, confirming that the LLM-generated problems are scientifically reasonable and representative of genuine materials simulation workflows, albeit with occasional AI artifacts.

Our study provides several insights with broad implications for the role of LLMs in materials science research. 
\noindent 
\begin{enumerate}[nosep]
     \item General-purpose LLMs consistently outperform domain-specific models in applying materials simulation tools, suggesting that broad knowledge coverage and reasoning ability are currently more critical for navigating complex scientific software than narrowly focused training.
    \item We demonstrated a strong synergy between LLMs and scientific tool documentation. RAG systems grounded in LLM-generated documentation achieved substantial performance gains, highlighting a phenomenon we describe as ``AI knows AI'', where models are more adept at interpreting and utilizing information generated by another AI.  
This suggests promising directions for designing AI- and user-friendly scientific software interfaces by leveraging AI-generated resources.
    \item Our results show that streamlined agent systems tend to outperform more complex designs.  
A self-reflection mechanism, combined with a simple RAG agent using LLM-generated documentation, surpassed more sophisticated multi-agent approaches such as Agentic RAG and GraphRAG.  
This ``simpler is better'' outcome challenges the assumption that increasing architectural complexity necessarily leads to better performance, and instead emphasizes the importance of refining retrieval sources and feedback loops.
\end{enumerate}

Error analysis  revealed that as models and systems improve, failure modes shift from surface-level issues (e.g., syntax errors, API misuse) to deeper logical challenges. 
For example, our self-reflection agent nearly eliminated import and API-related errors but exhibited concentrated failures in scientific logic. 
This shift from the ``how'' of code generation to the ``what'' of scientific reasoning underscores a critical frontier: enhancing the logical reasoning abilities of LLMs for scientific applications.
Overall, MatTools establishes a standardized framework for benchmarking and advancing LLM-tool interactions in materials science. 
The insights from this work illuminate the current landscape of LLMs in materials science and offer a clear path toward developing more intelligent and capable AI-assisted systems. 
Beyond evaluation, it provides QA datasets and real-world tool-usage examples that can serve as valuable resources for fine-tuning LLMs toward scientific coding tasks. 
The demonstrated advantages of general-purpose models, AI-generated documentation, and reflective agent systems point toward effective strategies for accelerating the integration of LLMs into computational materials science workflows and, ultimately, improving materials engineering.

Nonetheless, our approach has limitations. 
Expert review remains necessary to ensure scientific validity, as LLMs may generate physically inaccurate or non-standard descriptions. 
The modest scale of the current benchmark limits statistical robustness, though repeated trials mitigate this effect. 
Furthermore, the high computational cost of large-scale evaluations may restrict broader adoption. 


\section{Methods}\label{methods}
\subsection{Code extraction and document generation}
\begin{figure}[h]
    \centering
    \includegraphics[width=\textwidth]{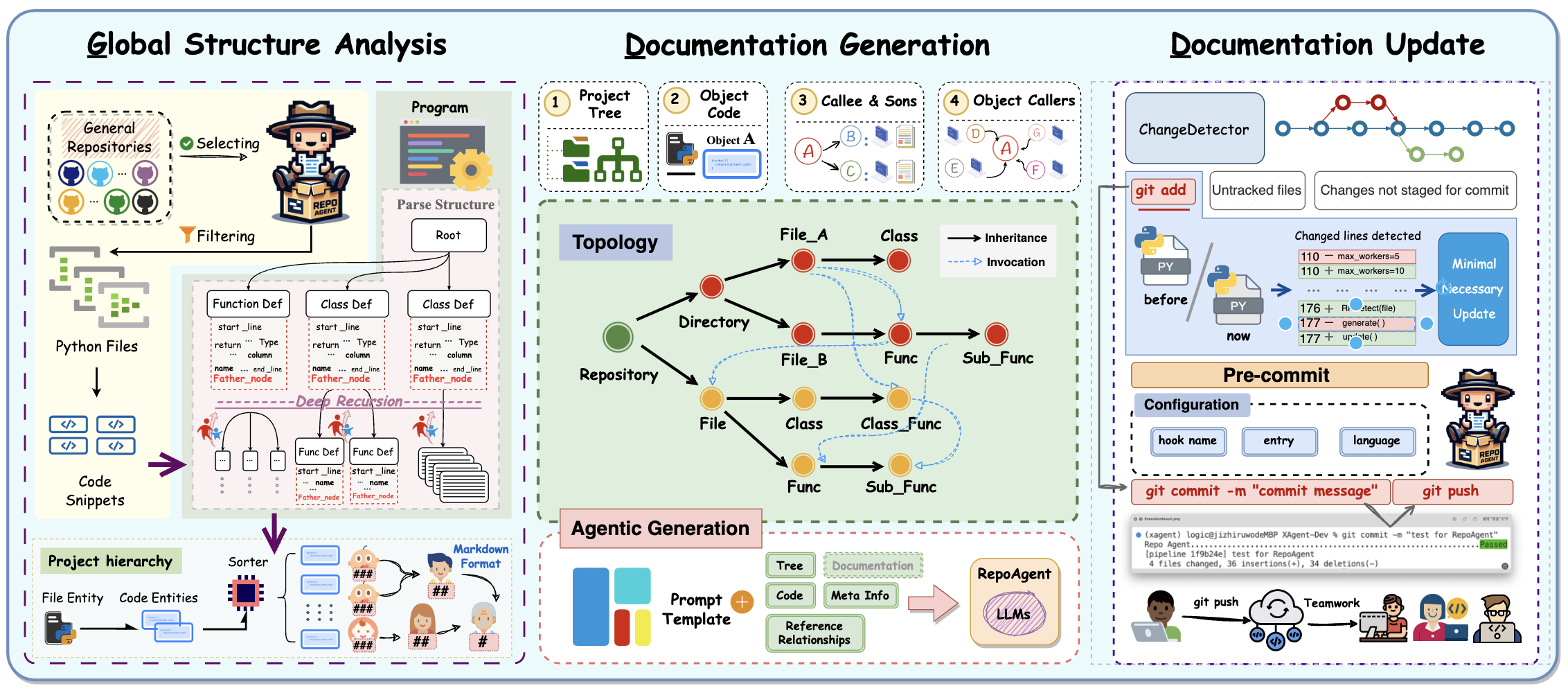}
    \caption{Three steps of \texttt{RepoAgent}: global structure analysis, documentation generation, and documentation update. This figure is from~\citep{luo-etal-2024-repoagent}. We used the first two steps of \texttt{RepoAgent} to extract the code segments and generate the documentation for the pymatgen software.}\label{fig:repoagent_method}
\end{figure}

\begin{figure}[h]
    \centering
    \includegraphics[width=0.4\textwidth]{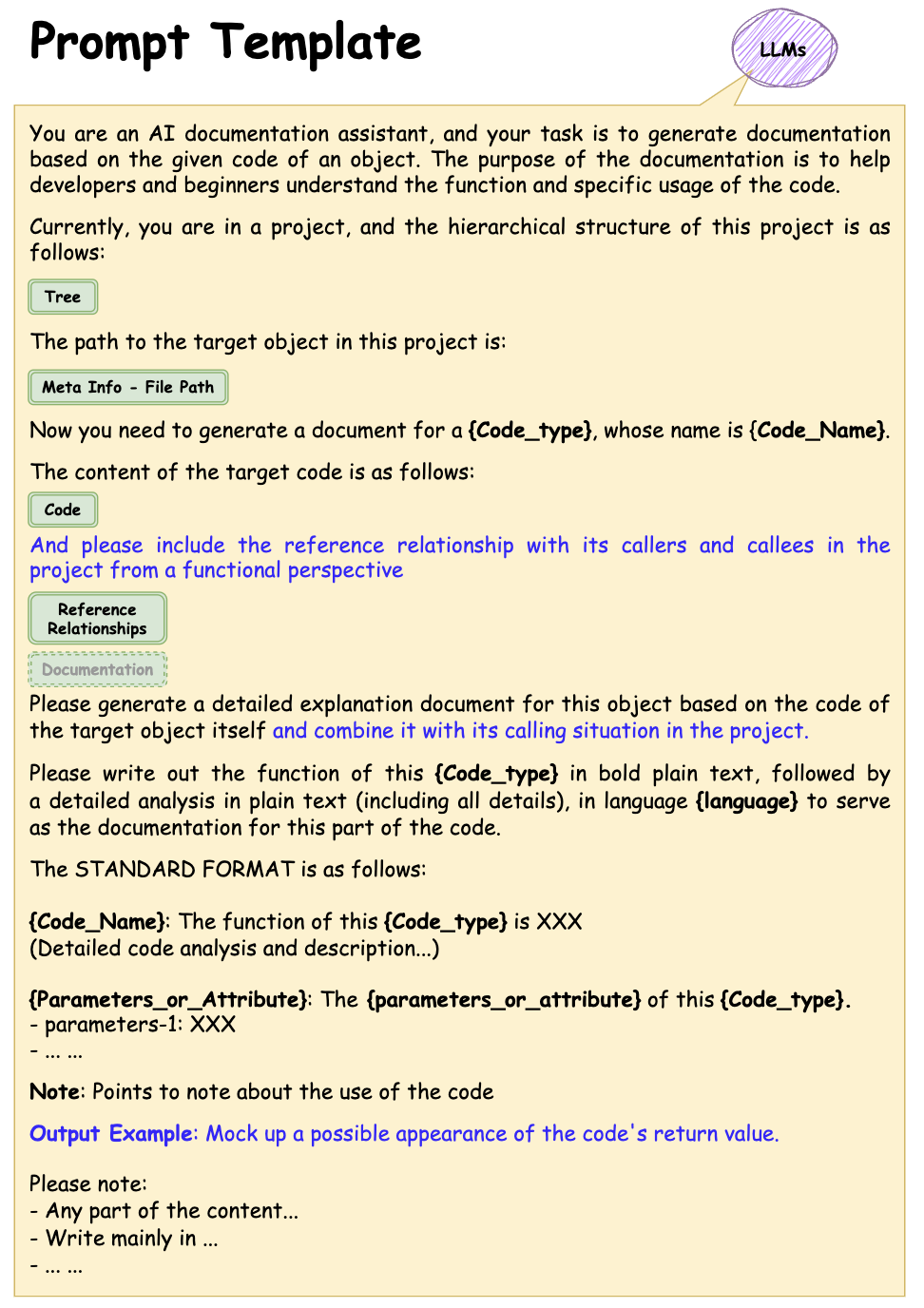}
    \caption{Prompt template of \texttt{RepoAgent}. This figure is from~\citep{luo-etal-2024-repoagent}.}\label{fig:repoagent_prompt_template}
\end{figure}

We used \texttt{RepoAgent} to extract the code segments and generate the documentation for the \texttt{pymatgen} software. We keep the same prompt template (Supplementary Fig.~\ref{fig:repoagent_prompt_template}) and settings as \texttt{RepoAgent} for the global structure analysis and documentation generation (Supplementary Fig.~\ref{fig:repoagent_method}). The model used to generate the documentation is \texttt{Gemini-2.0-flash}. The example of extracted code segment and generated documentation is shown in Supplementary Fig.~\ref{fig:code_segment_and_generated_doc}.

\subsection{Settings for materials simulation tool QA benchmark}
The versions used in our codebase are 2024.8.9 for pymatgen and 2024.7.19 for pymatgen-analysis-defects. All tests were conducted on an NVIDIA A100 80GB GPU. For both QA generation and testing, we maintained a consistent temperature setting of 0.7 across all models while keeping other parameters at their default values in the OpenAI API client. We verified all generated QA pairs for proper formatting and content completeness, ensuring each pair contained a question, four answer choices, and the correct answer. For the Gemini series models, we utilized the Gemini API in conjunction with the OpenAI API client. For the Qwen series and materials chemistry models, we deployed vllm to emulate the OpenAI API client interface.

\subsection{Synthesis of real-world tool-usage benchmark}
Examples of real-world materials simulation tool usage are scarce, making it challenging to directly construct evaulation datasets. Hence, we developed an automated pipeline using LLMs to transform materials property test code into triplets of: (1) problem statement: A natural-language prompt that instructs the LLMs to generate Python code for calculating material properties and returning a dictionary of results, (2) expected property dictionary: A set of key-value pairs representing the material property names, their expected values, and data types, which serve as the reference for verification, and (3) verification code: Scripts that automatically validate whether the generated results match the expected outputs in (2). We selected material property test code as the source because it inherently contains all three essential components: the scientific problem, the implementation of a solution, and result verification. 
This automated pipeline enables the systematic and scalable construction of tool-usage datasets, independent of any specific LLM, thereby facilitating fair benchmarking across models.

In practice, we curated test files from the \texttt{pymatgen-analysis-defects} library, a standalone \texttt{pymatgen} plugin developed for defect analysis (important material properties are controlled by the defects in materials). 
We first split the test files into individual Python functions, from which GPT-4o~\citep{gpt4o} automatically generated triplets. Two PhD students in materials science manually reviewed and corrected errors in the generated triplets.
We generated \textbf{49 questions} (\textbf{138 tasks}, where each task corresponds to the calculation of a specific material property) for real-world tool-usage benchmark. 
We also invited independent materials science experts to evaluate the quality of generated questions. (See Supplementary Section~\ref{sec:a2.1} for triplet generation prompts, representative examples and human review and evaluation procedure.) 

\subsection{Docker sandbox for result checking}
We implemented a Docker-based sandbox environment to ensure safe and reproducible evaluation of LLM-generated code. 
This sandbox isolates code execution from the local system and provides two key functionalities: (1) executing the LLM-generated Python code and returning its output in the form of a materials property dictionary, and (2) executing the corresponding verification scripts to assess correctness. The verification returns ``ok'' when the results match the expected outputs, and an error list otherwise.

\subsection{Knowledge cut-off date of LLMs}
All evaluated LLMs have knowledge cut-off dates that precede or approximately coincide with the release dates of the libraries used in our benchmark (August 9, 2024, for \texttt{pymatgen} and July 19, 2024, for \texttt{pymatgen-analysis-defects}). 
Given the generally low accuracy of standalone LLMs in our real-world tool-usage benchmark, we can reasonably exclude the possibility of data leakage from these libraries into the training corpora of the tested models.

\subsection{\textbf{Answer extraction from LLM responses}} we implemented a parsing mechanism that identifies content enclosed within \verb|<answer>| and \verb|</answer>| tags. This approach ensures standardized answer extraction and evaluation. If a model failed to adhere to the required format (see Supplementary Fig.~\ref{fig:qa_test_prompt}) in its response, we considered it as a failed attempt and marked the answer as incorrect.

\section*{Data availability}
Datasets: \url{https://github.com/Grenzlinie/MatTools}

\section*{Code availability} 
Source code: \url{https://github.com/Grenzlinie/MatTools}

\def\bibsection{\section*{\refname}}
\bibliography{./manuscript.bib}

\section*{Declarations}
\textbf{Funding}
The authors would like to thank for startup funding from Materials Innovation Institute for Life Sciences and Energy (MILES), HKU-SIRI in Shenzhen for support of this manuscript. This work is supported by Research Grants Council, Hong Kong SAR through the General Research Fund (17210723, 17200424). T. W. acknowledges additional support by The University of Hong Kong (HKU) via seed funds (2509100468) and Guangdong Natural Science Fund (2025A1515012129).

\section*{Author contributions}
S.L. designed and implemented the materials simulation tool QA benchmark and the real-world tool-usage benchmark, conducted experiments, analyzed results, and wrote the initial manuscript draft. B.H. and B.Y. helped conduct the human verification test. J.X. helped implement the materials simulation tool QA benchmark, conducted experiments, and analyzed results. B.Y. and B.H. assisted in data curation and result analysis. D.J.S. and T.W. supervised the project, provided critical feedback on the study design and results interpretation, and contributed to manuscript revisions. All authors reviewed and approved the final manuscript.

\section*{Competing interests}
The authors declare no competing interests.

\section*{Supplementary Informations}
Supplementary Figures S1-S17

Supplementary Tables S1-S8

\clearpage
\pagebreak
\beginsupplement
\pagestyle{plain}
\onecolumngrid

\begin{center}
  \huge {Supplementary Information for}
  \bigskip

  \large \textbf{MatTools: Benchmarking Large Language Models for Materials Science Tools}

  \bigskip
  \bigskip

Siyu Liu, Bo Hu, Beilin Ye, Jiamin Xu, David J. Srolovitz and Tongqi Wen
\end{center}

\clearpage

\makeatletter

\titleformat{\subsubsection}
{\large}{\thesubsubsection}{1em}{{#1}}

\titleformat{\subsection}
{\bfseries\large}{\thesubsection}{1em}{{#1}}

\titleformat{\section}
{\bfseries\Large}{\thesection}{1em}{{#1}}

\section{Details of materials simulation tool QA benchmark and results}\label{sec:a1}

\subsection{Method, prompt and generated document example of RepoAgent}\label{sec:a1.1}

\FloatBarrier

\FloatBarrier

\begin{figure}[h]
    \centering
    \includegraphics[width=\textwidth]{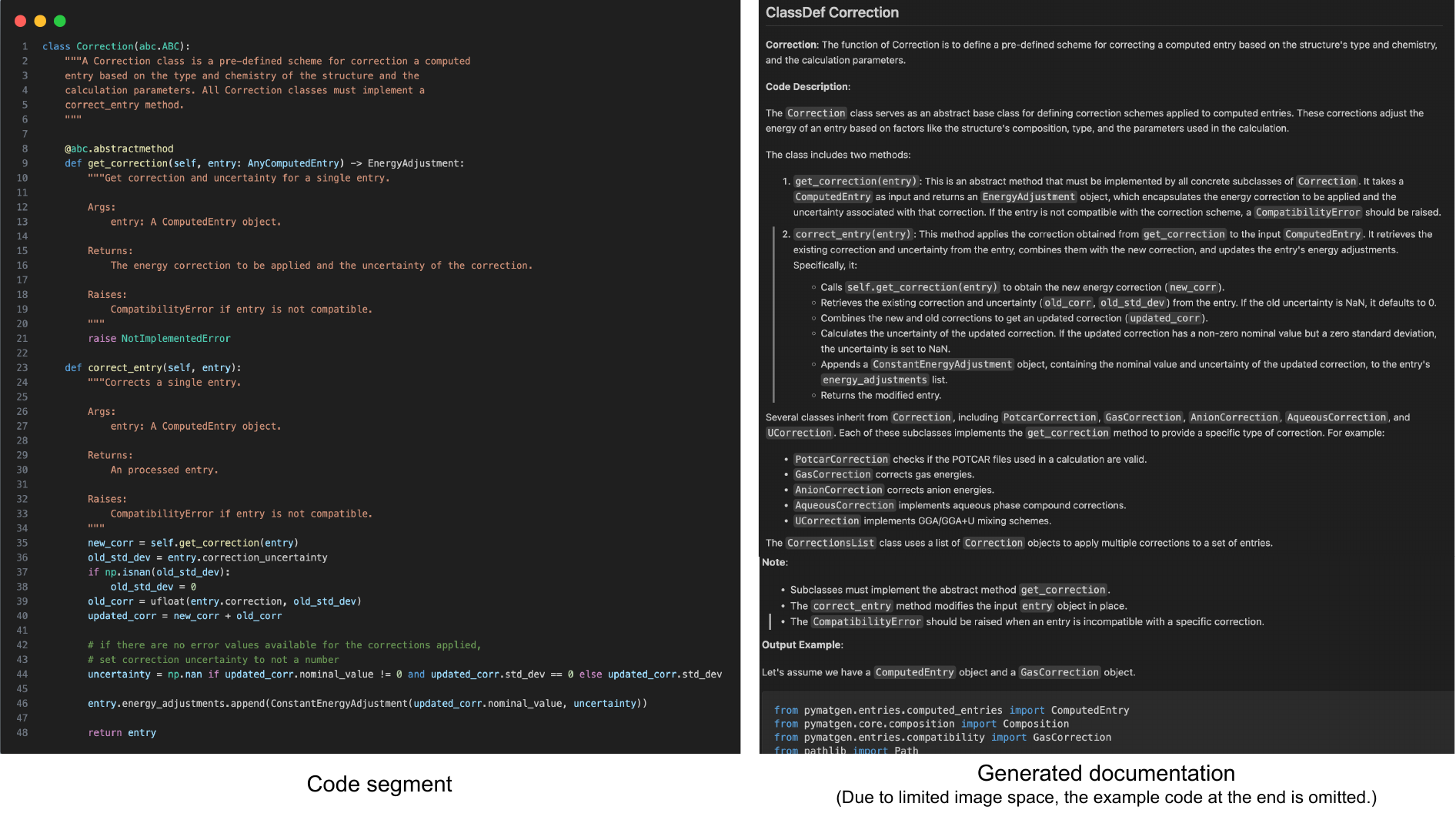}
    \caption{Example of extracted code segment and generated documentation for \texttt{pymatgen}.}\label{fig:code_segment_and_generated_doc}
\end{figure}
\FloatBarrier

\newpage
\subsection{QA benchmark prompt templates}\label{sec:a1.2}

\begin{figure}[h]
    \centering
    \includegraphics[width=0.6\textwidth]{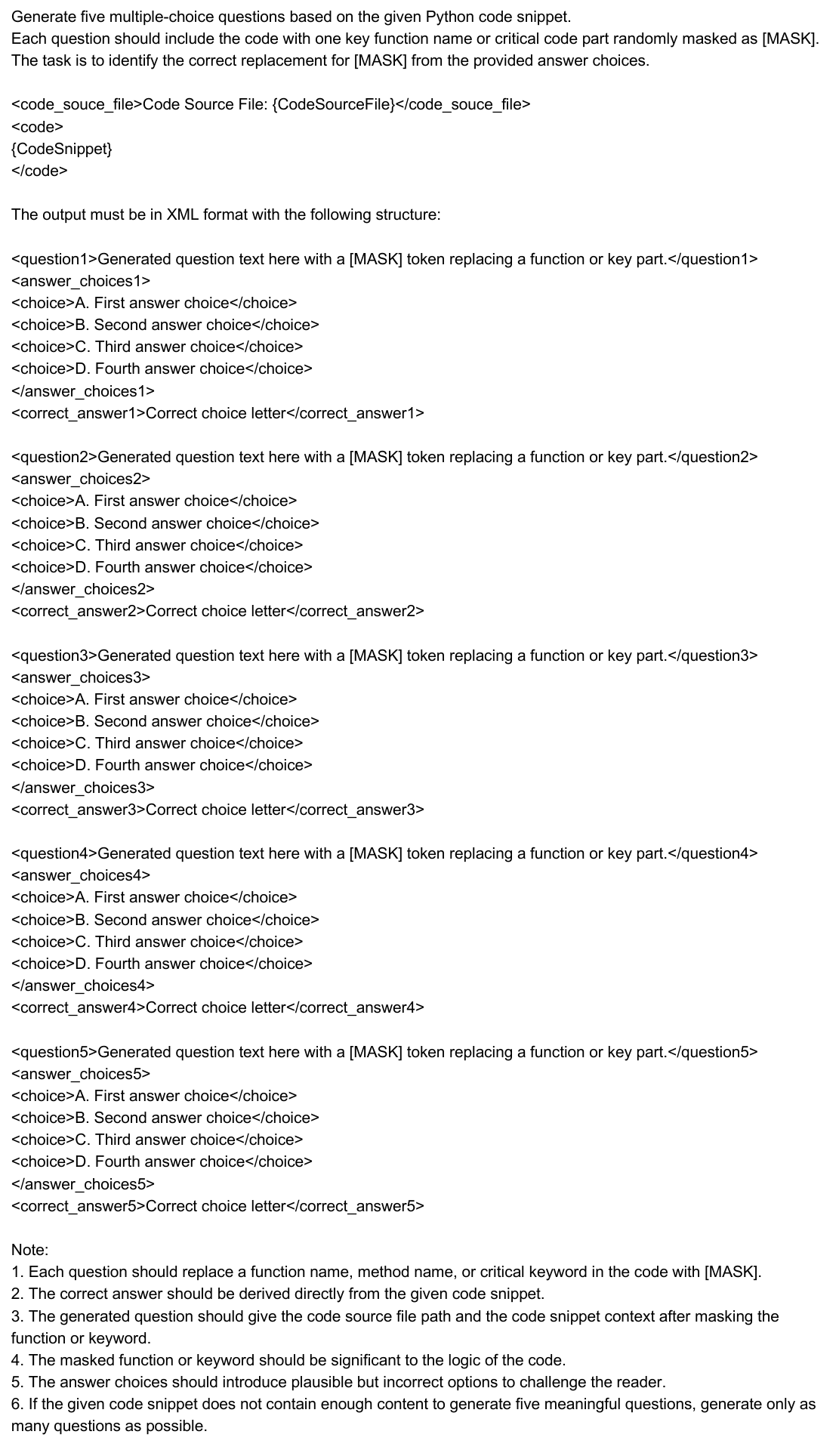}
    \caption{Prompt template for generating the QA pairs for \texttt{pymatgen} code.}\label{fig:code_qa_prompt}
\end{figure}
\FloatBarrier

\begin{figure}[h]
    \centering
    \includegraphics[width=0.6\textwidth]{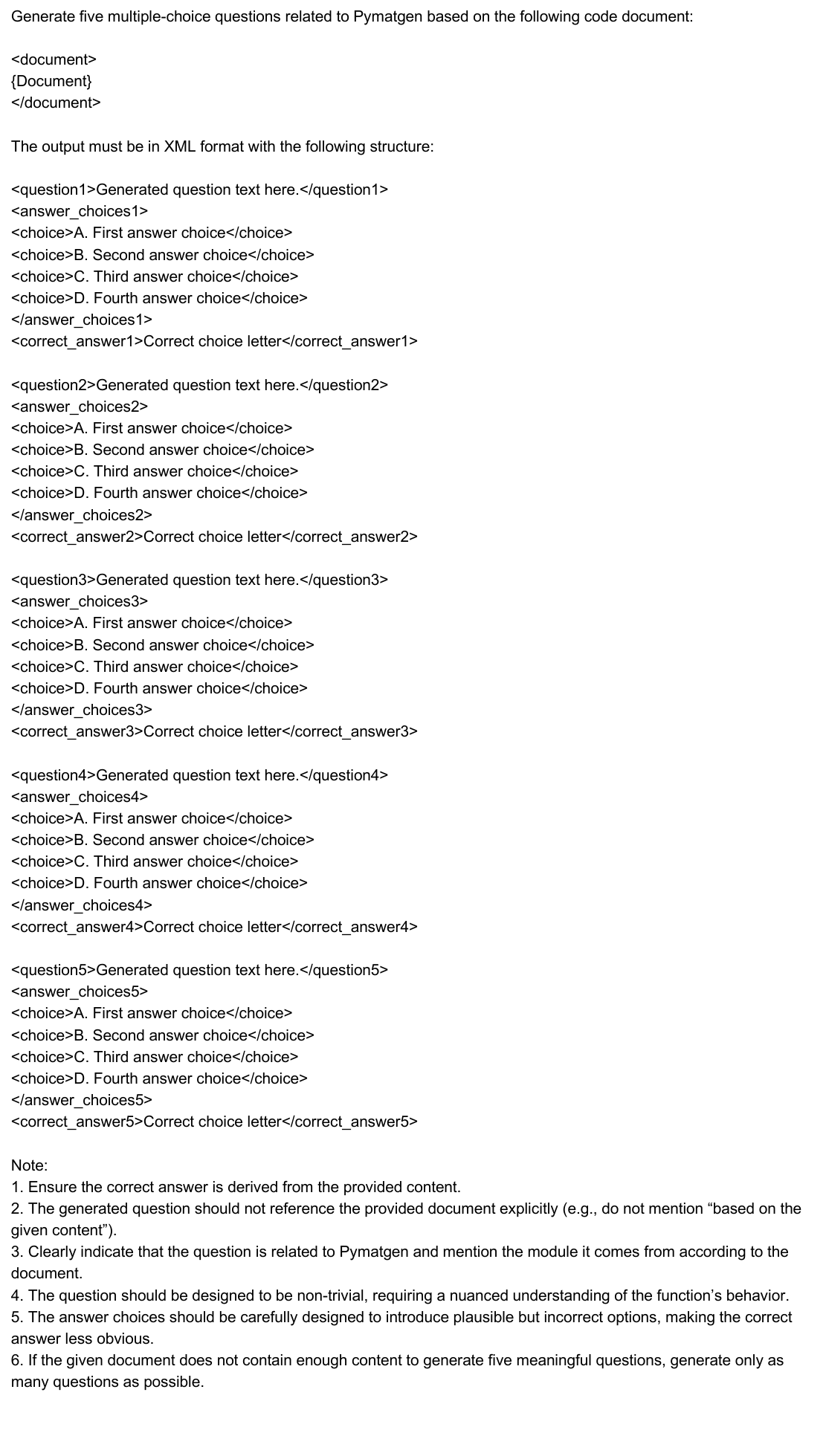}
    \caption{Prompt template for generating the QA pairs of \texttt{pymatgen} documentation.}\label{fig:doc_qa_prompt}
\end{figure}
\FloatBarrier

\begin{figure}[h]
    \centering
    \includegraphics[width=0.8\textwidth]{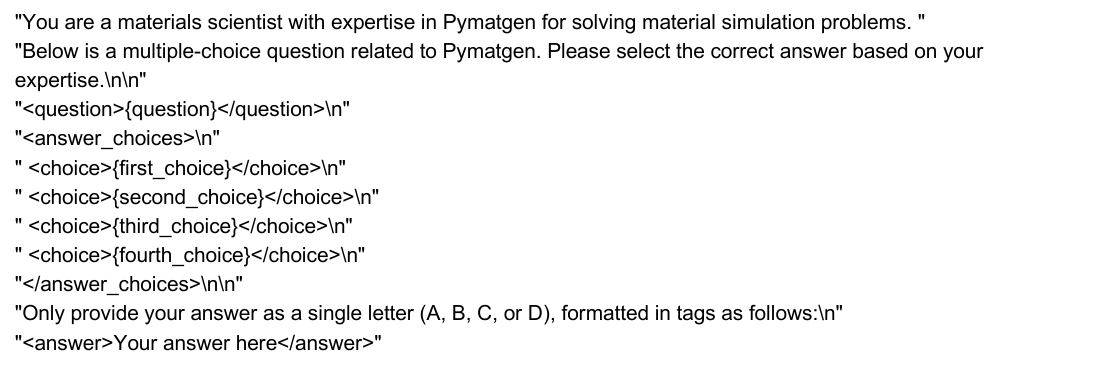}
    \caption{Prompt template for instructing the LLM to answer questions in the QA benchmark.}\label{fig:qa_test_prompt}
\end{figure}
\FloatBarrier

\newpage
\subsection{Examples of generated QA pairs and testing results}\label{sec:a1.3}

\begin{figure}[htb]
    \centering
    \includegraphics[width=0.8\textwidth]{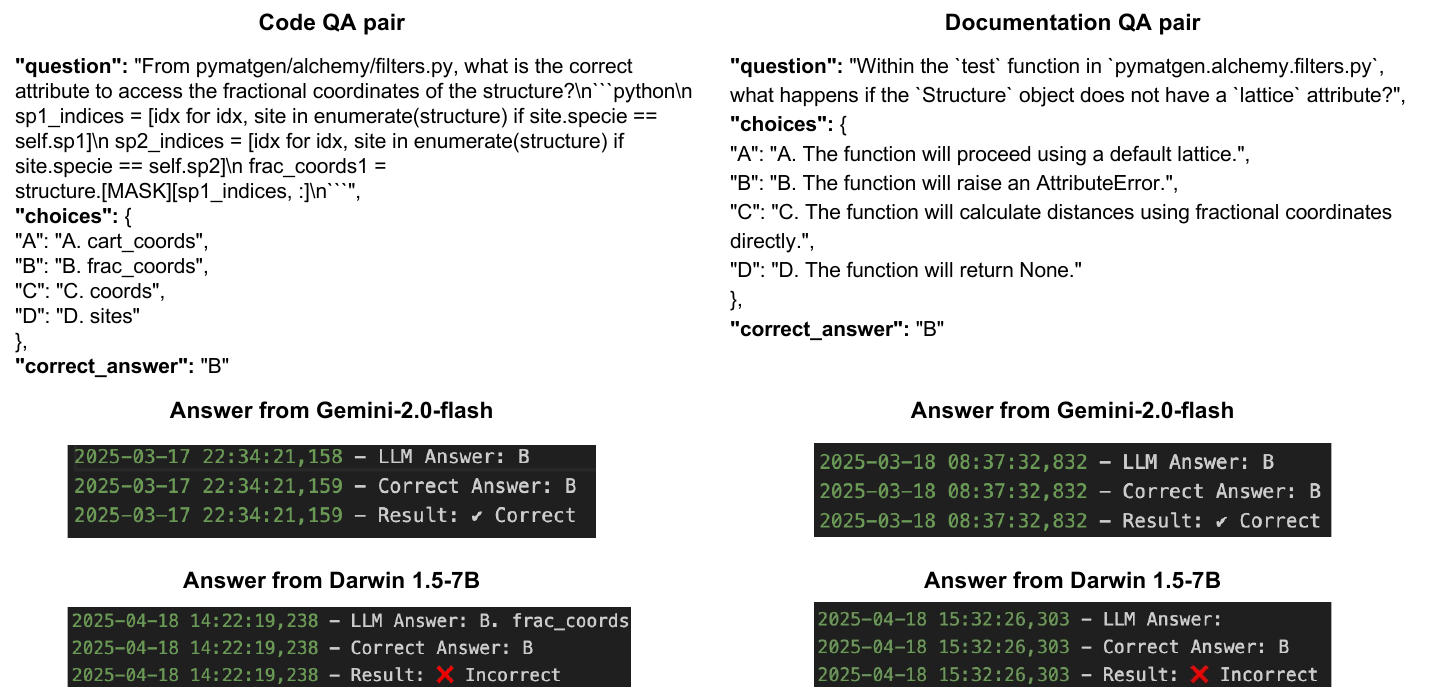}
    \caption{One example of generated QA pairs and testing results from different models. The Darwin 1.5-7B model failed to follow the required format in its response, so we considered it as a failed attempt and marked the answer as incorrect.}
    \label{fig:qa_results1}
\end{figure}

\begin{figure}[htb]
    \centering
    \includegraphics[width=0.9\textwidth]{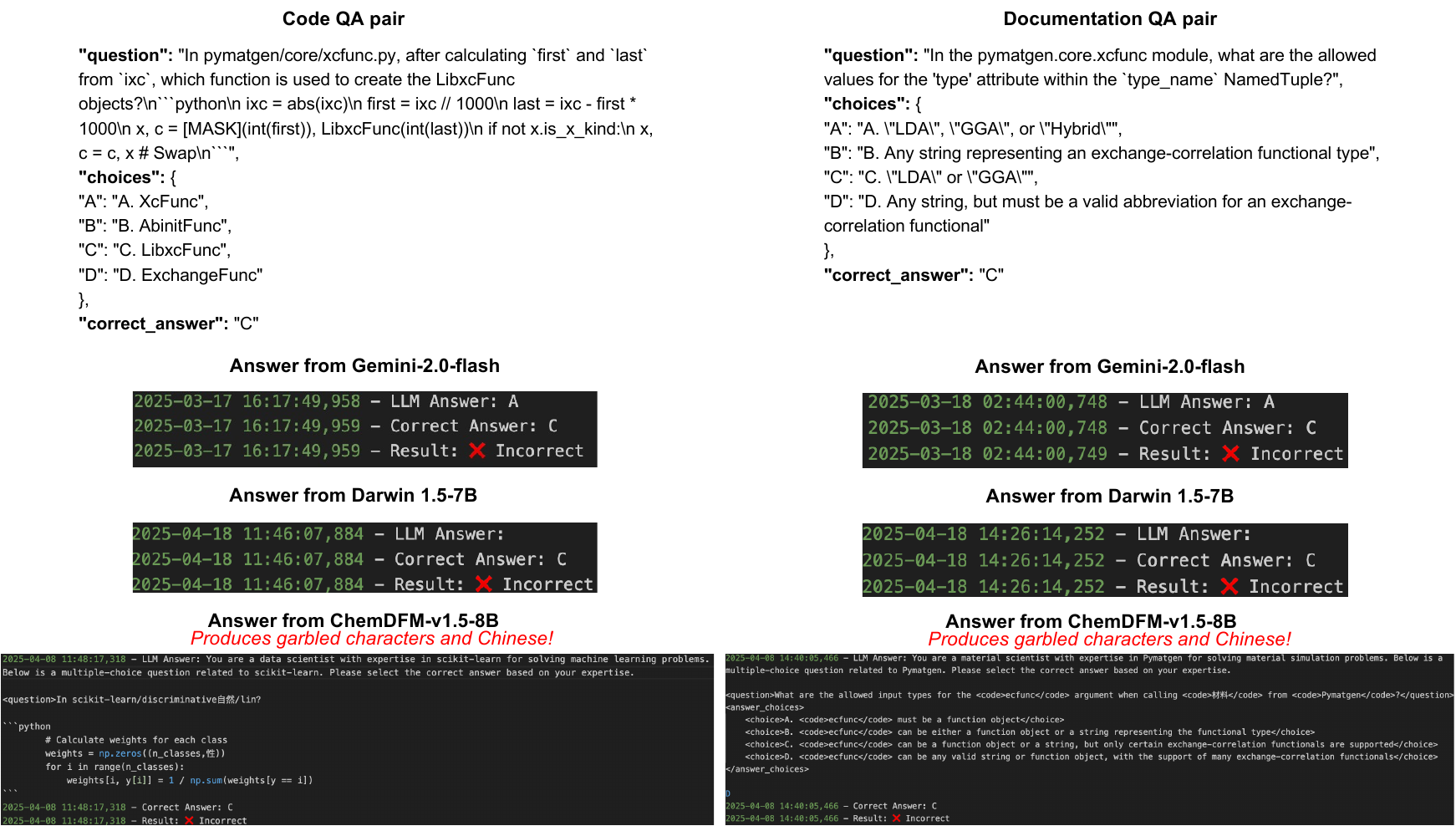}
    \caption{One example of generated QA pairs and testing results from different models. Both the closed-source and materials chemistry LLMs failed to output correct answers. The ChemDFM-v1.5-8B model even generated garbled characters and Chinese text.}\label{fig:qa_results2}
\end{figure}

\begin{verbatim}
extracted_answer = llm_answer.strip().replace("<answer>", "")
                    .replace("</answer>", "").strip()
\end{verbatim}
Here are some examples (Fig.~\ref{fig:qa_results1} and~\ref{fig:qa_results2}) of generated QA pairs and testing results from different models.

\FloatBarrier

\FloatBarrier

\subsection{Model list and settings for materials simulation tool QA benchmark}\label{sec:a1.4}

\paragraph*{\textbf{Model list}}
\begin{itemize}
    \item Gemini-1.5-Flash: use API with model name \texttt{gemini-1.5-flash}.
    \item Gemini-1.5-Pro: use API with model name \texttt{gemini-1.5-pro}.
    \item Gemini-2.0-Flash: use API with model name \texttt{gemini-2.0-flash}.
    \item Qwen2.5-7B-Instruct: use local vllm~\citep{kwon2023efficient} server and download the model from \url{https://huggingface.co/Qwen/Qwen2.5-7B-Instruct}.
    \item Qwen2.5-14B-Instruct: use local vllm server and download the model from \url{https://huggingface.co/Qwen/Qwen2.5-14B-Instruct}.
    \item Qwen2.5-Coder-14B-Instruct: use local vllm server and download the model from \url{https://huggingface.co/Qwen/Qwen2.5-Coder-14B-Instruct}.
    \item Qwen2.5-32B-Instruct: use local vllm server and download the model from \url{https://huggingface.co/Qwen/Qwen2.5-32B-Instruct}.
    \item Qwen2.5-Coder-32B-Instruct: use local vllm server and download the model from \url{https://huggingface.co/Qwen/Qwen2.5-Coder-32B-Instruct}.
    \item Qwen2.5-72B-Instruct: use local vllm server and download the model from \url{https://huggingface.co/Qwen/Qwen2.5-72B-Instruct-GPTQ-Int4}. Here we used the \texttt{gptq} version and \texttt{int4} quantization.
    \item ChemDFM-v1.5-8B: use local vllm server and download the model from \url{https://huggingface.co/OpenDFM/ChemDFM-v1.5-8B}.
    \item ChemLLM-7B-Chat-1\_5-DPO: use local vllm server and download the model from \url{https://huggingface.co/AI4Chem/ChemLLM-7B-Chat-1_5-DPO}. Due to the slow running speed and low instruction-following ability, we only test 6705 code QA pairs and 32006 documentation QA pairs.
    \item Darwin 1.5-7B: use local vllm server and download the model from \url{https://github.com/MasterAI-EAM/Darwin}.
\end{itemize}

\begin{table}[htbp]
	\centering
	\caption{Accuracy of reasoning-enabled LLMs on the materials simulation tool QA benchmarks, evaluated on code-based (\texttt{pymatgen\_code\_qa}) and documentation-based (\texttt{pymatgen\_doc\_qa}) tasks.}
	\begin{tabular}{ccc}
		\toprule
		\textbf{Model} & \textbf{Accuracy (\%)} & \textbf{Question number and type} \\
		\midrule
		qwq-32b      & 84.27 & 34621 (full code QA pairs) \\
		qwen3-4b     & 78.3  & 493 (code) \\
		qwen3-8b     & 81.75 & 674 (code) \\
		qwq-32b      & 84.12 & 422 (doc) \\
		qwen3-4b     & 76.16 & 537 (doc) \\
		qwen3-8b     & 78.58 & 691 (doc) \\
		\bottomrule
	\end{tabular}\label{tab:reasoning_qa_results}
\end{table}
\FloatBarrier

\newpage
\section{Details for real-world tool-usage benchmark and results}\label{sec:a2}

\subsection{Example prompt and generated triplet for the benchmark evaluation}\label{sec:a2.1}

\begin{figure}[htb]
    \centering
    \includegraphics[width=0.8\textwidth]{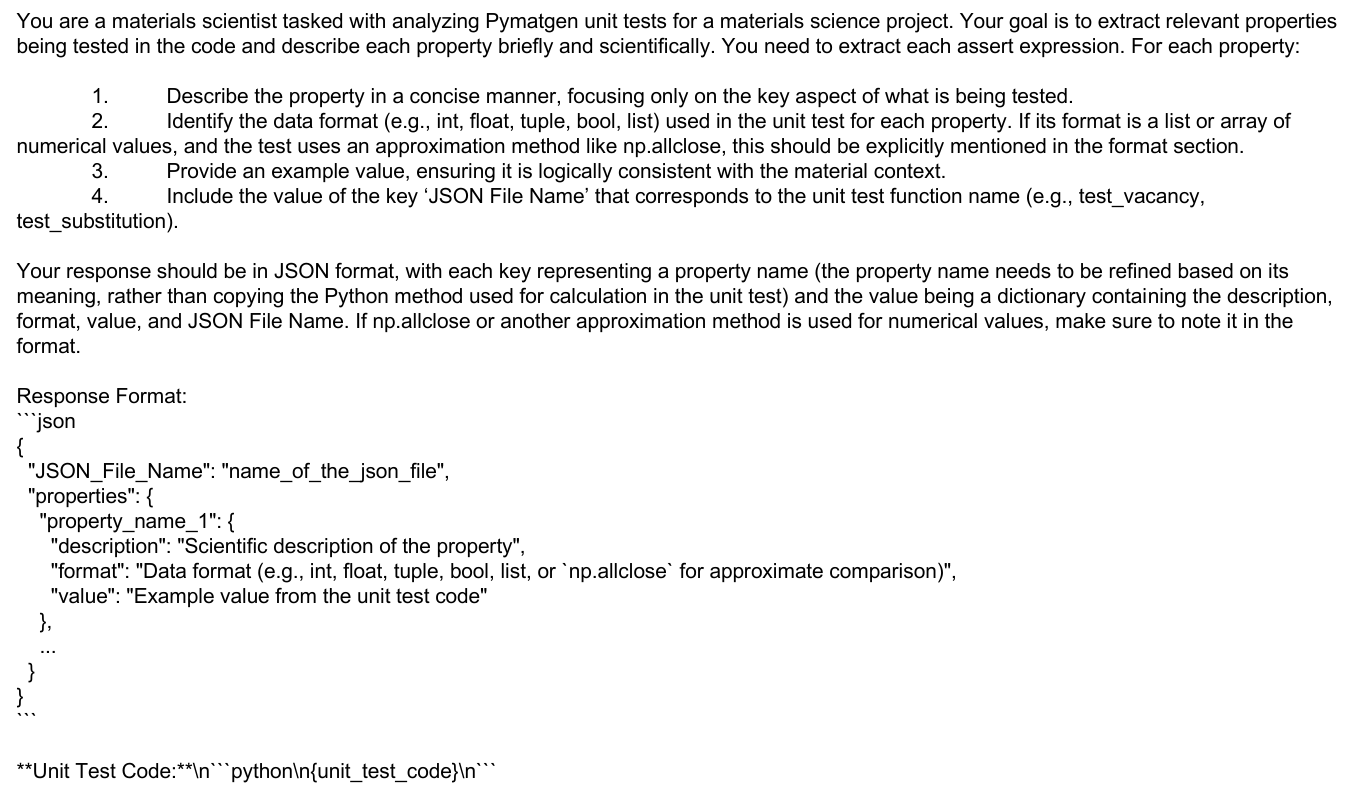}
    \caption{Prompt template for extracting the materials properties from the unit test code.}\label{fig:extract_property_prompt}
\end{figure}
\FloatBarrier

\begin{figure}[htb]
    \centering
    \includegraphics[width=0.9\textwidth]{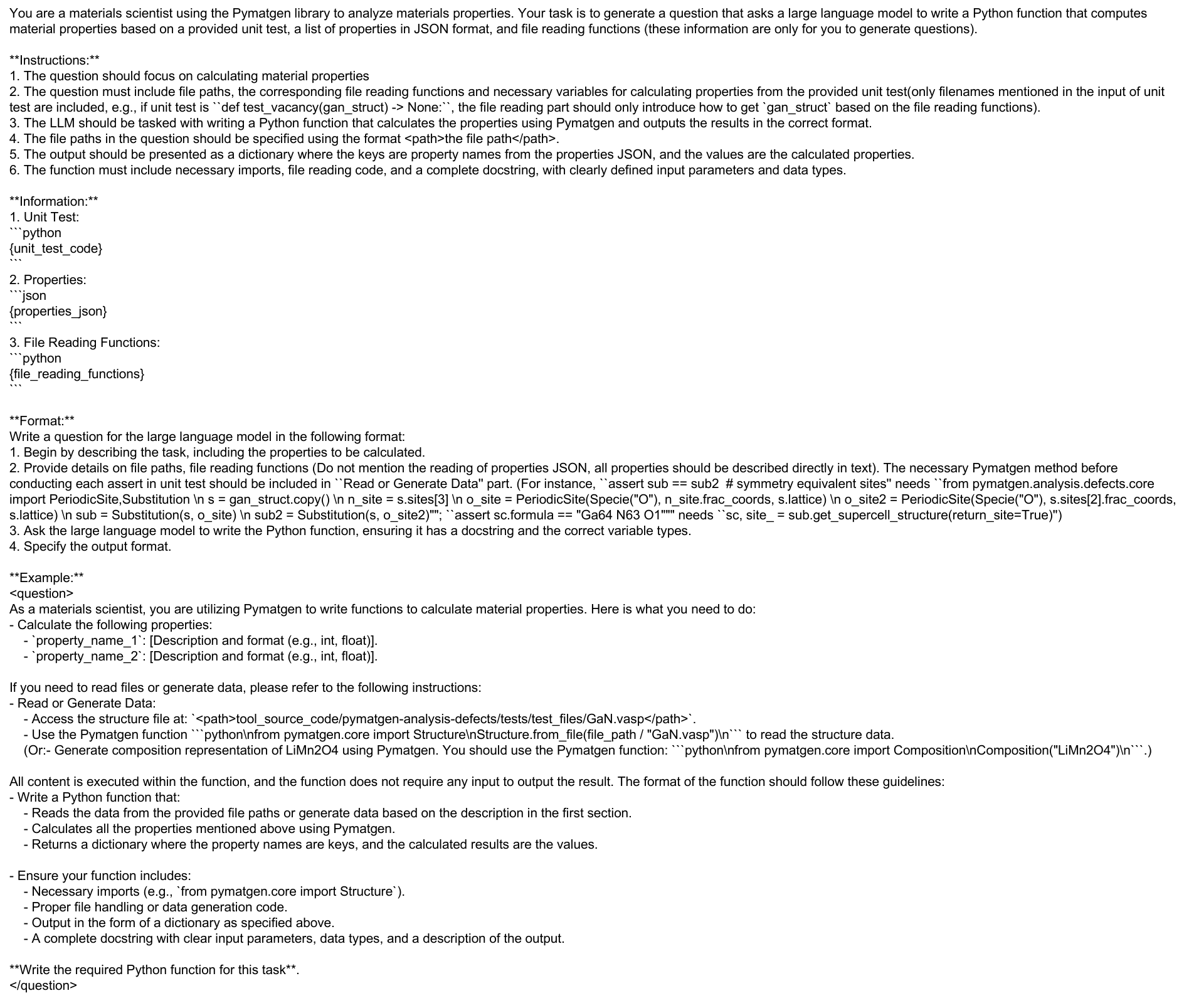}
    \caption{Prompt template for proposing the problem statement from the unit test code.}\label{fig:build_question_prompt}
\end{figure}
\FloatBarrier

\begin{figure}[htb]
    \centering
    \includegraphics[width=\textwidth]{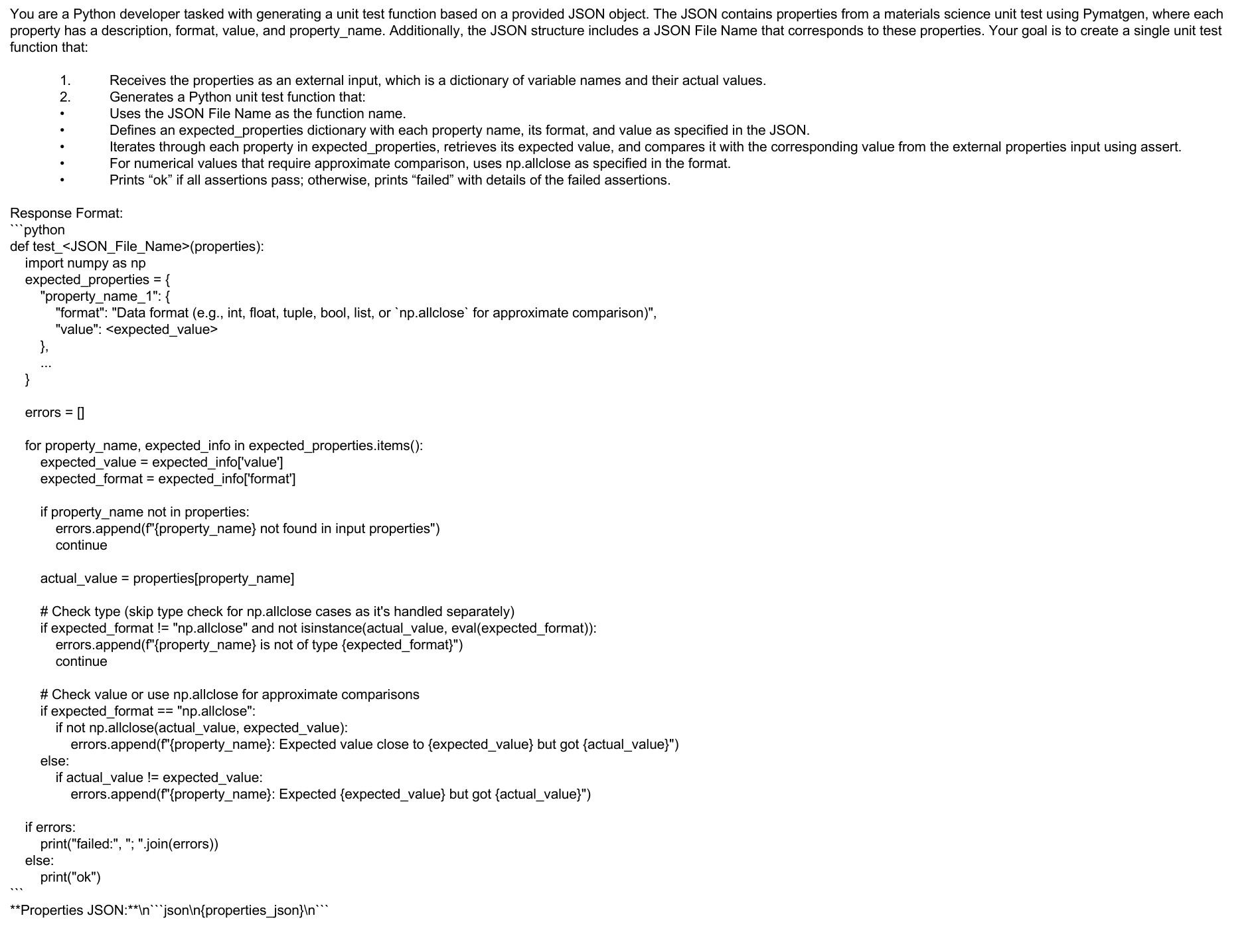}
    \caption{Prompt template for generating the verification code for the problem statement.}\label{fig:generate_unit_test_prompt}
\end{figure}
\FloatBarrier

\begin{figure}[htb]
    \centering
    \includegraphics[width=\textwidth]{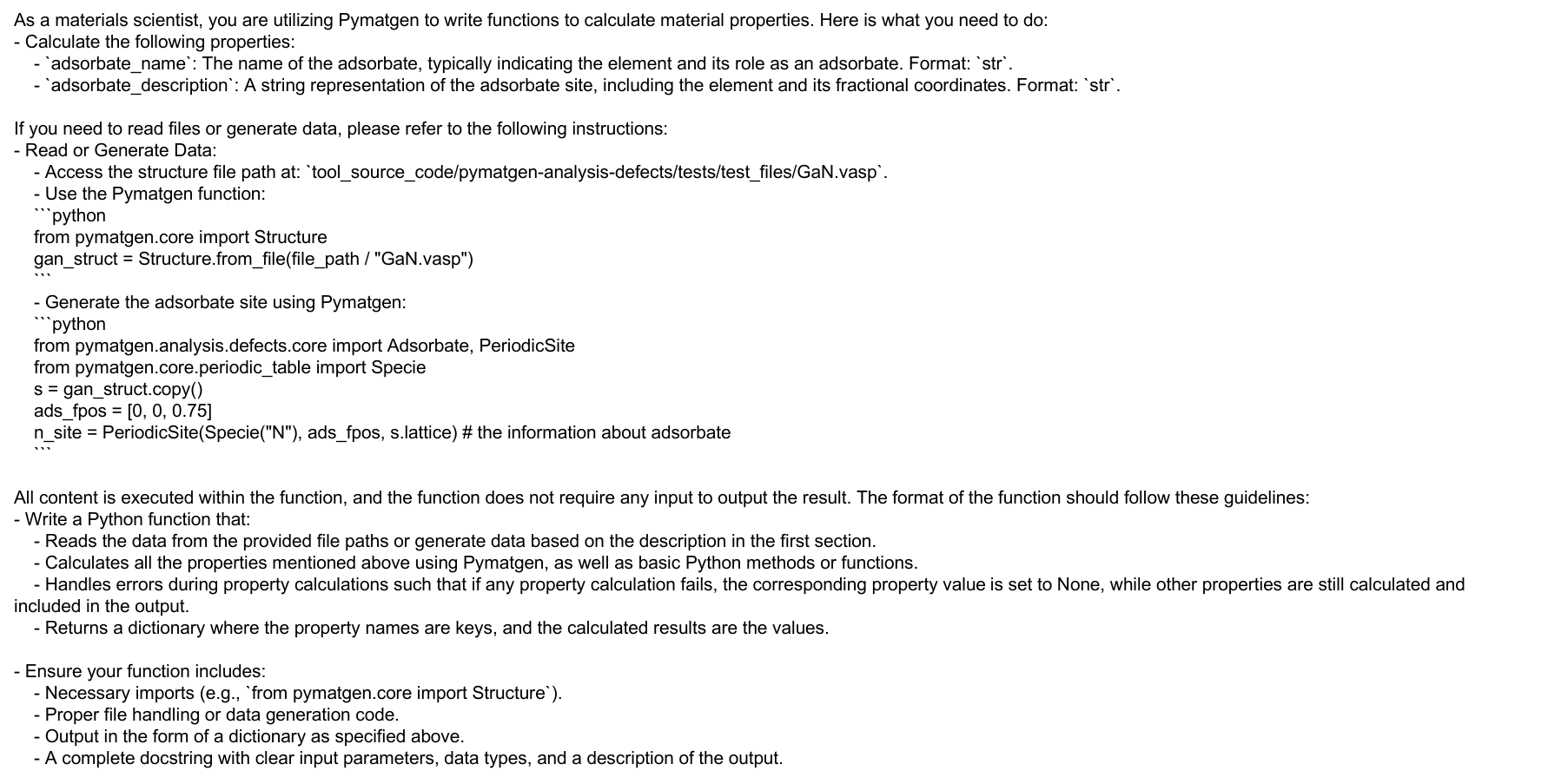}
    \caption{Example of the problem statement.}\label{fig:problem_statement_example}
\end{figure}
\FloatBarrier

\begin{figure}[htb]
    \centering
    \includegraphics[width=\textwidth]{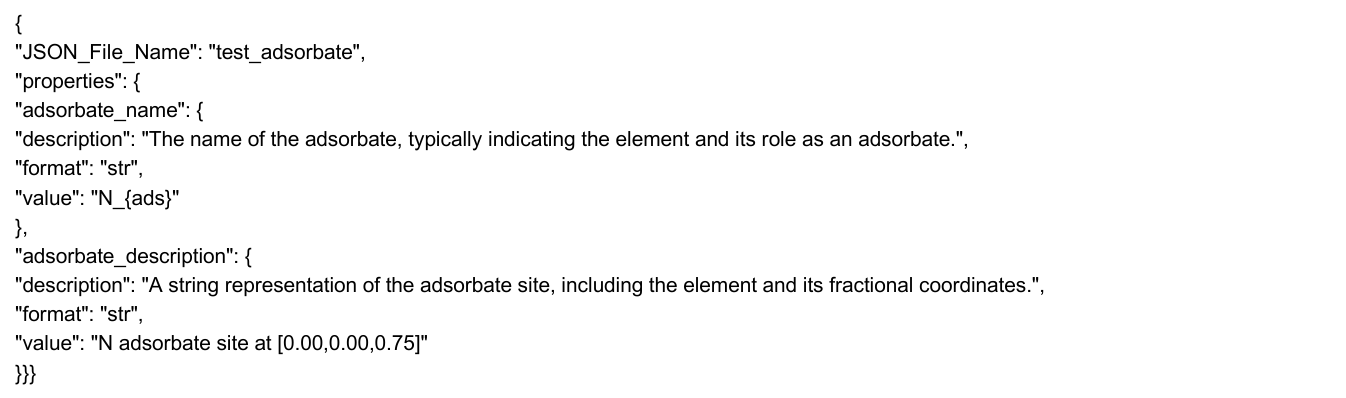}
    \caption{Example of the property dictionary.}\label{fig:property_dict_example}
\end{figure}
\FloatBarrier

\begin{figure}[htb]
    \centering
    \includegraphics[width=\textwidth]{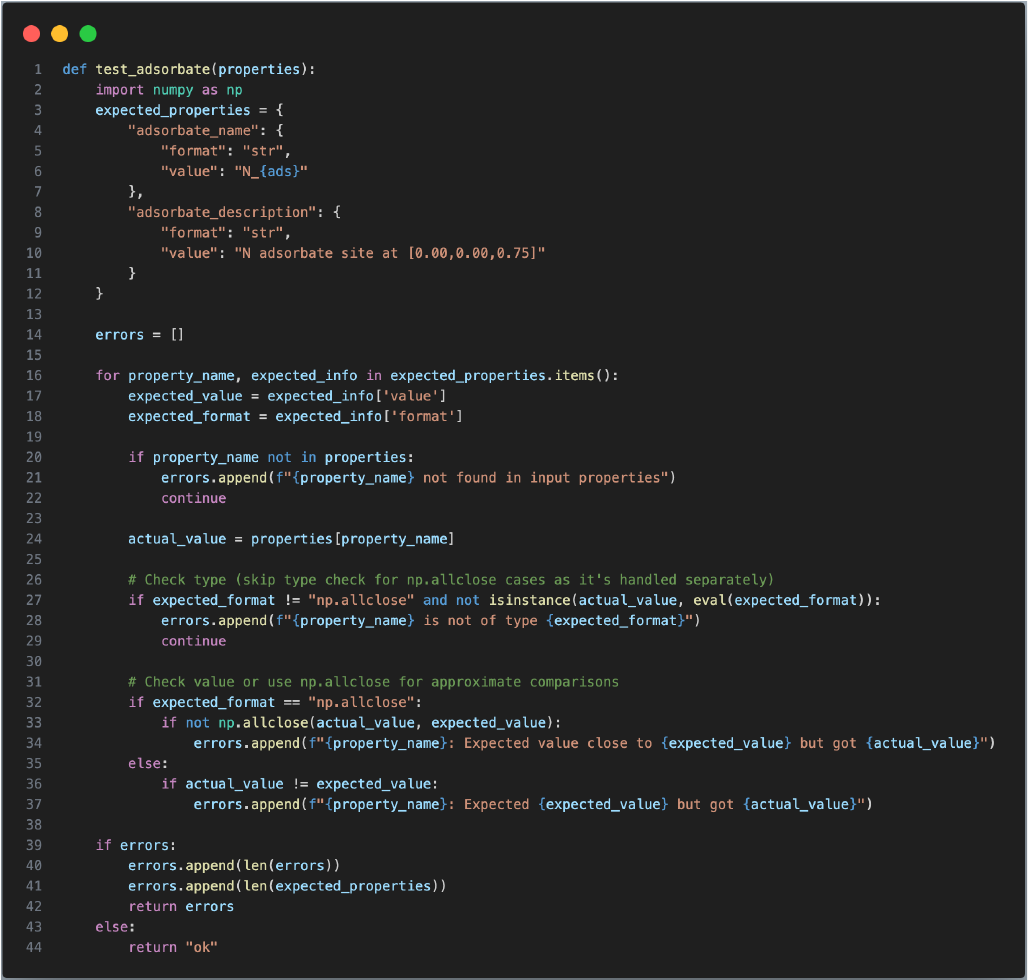}
    \caption{Example of the verification code.}\label{fig:verification_code_example}
\end{figure}
\FloatBarrier

\begin{figure}[htb]
    \centering
    \includegraphics[width=\textwidth]{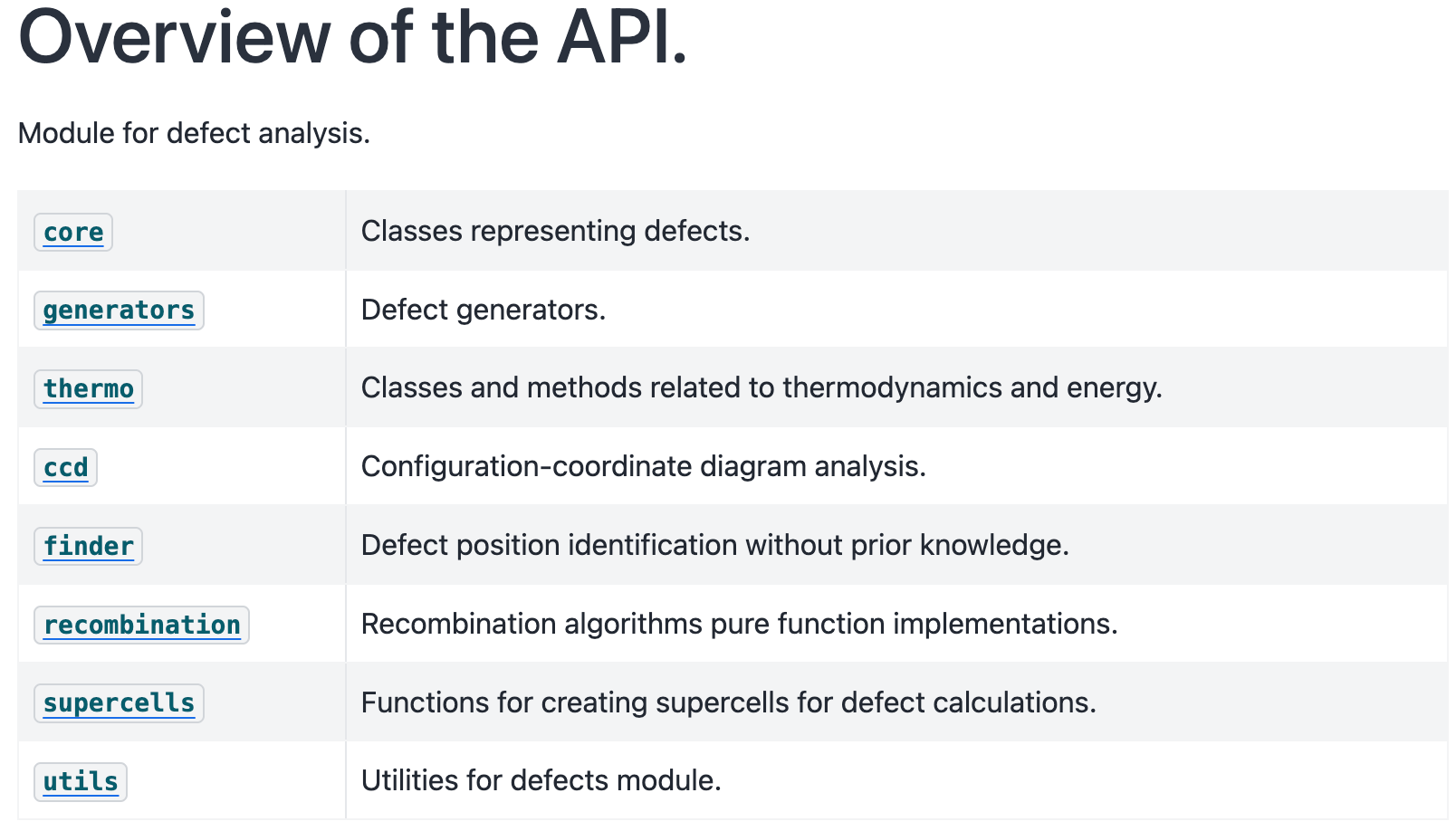}
    \caption{Overview of the functions of \texttt{pymatgen-analysis-defects}~\citep{pymatgen-defects} package.}\label{fig:pymatgen_api_overview}
\end{figure}
\FloatBarrier

\newpage
\paragraph*{\textbf{Benchmark scope derived from \texttt{pymatgen-analysis-defects} package}}
Our benchmark is derived from the specific classes and functions of these APIs (Fig.~\ref{fig:pymatgen_api_overview}).

\paragraph*{\textbf{Human review process of the triplets}}
We conducted a human-in-the-loop evaluation to ensure the quality and scientific validity of the real-world tool-usage benchmark. 
As the paper states, two materials science PhD students reviewed and revised the generated triplets. 
This process involved the following systematic steps:

1.	Independent Review: Each of the 49 generated task triplets was independently reviewed by both PhD students. A triplet consists of (1) a problem statement, (2) a dictionary of expected material properties, and (3) verification code.

2.	Revision Criteria: the reviewers assessed each triplet based on a consistent set of criteria:
\noindent 
\begin{itemize}[nosep]
    \item Scientific Validity: the reviewers verified that the physical descriptions of materials properties were scientifically accurate and reasonable.
    \item Correctness of Verification: the reviewers ensured that the verification code and the expected values for material properties were correct, as derived from the original material property test files.
    \item Clarity and Feasibility: the reviewers checked that the problem statement was clear, unambiguous.
    \item Format Adherence: the reviewers confirmed that all parts of the triplet adhered to the required data structures and formats necessary for the automated benchmark pipeline to function correctly.
\end{itemize}

3.	Inter-Reviewer Agreement and Discrepancy Resolution: after the independent review, the reviewers compared their findings. 
Discrepancies were resolved through a discussion to reach a consensus. 
If an agreement could not be reached on a generated task, or if it was deemed scientifically flawed or ambiguous, it was revised or discarded from the benchmark. 
This consensus-based approach ensures that every task in the final benchmark is vetted for quality and accuracy by multiple experts.

Of course, this human-in-the-loop approach is crude, as may be expected for a prototype study. 
Our human-in-the-loop process primarily focused on whether the description of the task triplet is complete and correct (i.e., meeting the format requirements set in our generation prompt), whether the generated property description is scientifically reasonable, and whether the standard answer to the material property test is correct.

\paragraph*{\textbf{Small study -- human evaluation of the quality of generated questions}}
Five independent materials science domain experts (A to E) were involved in the evaluation. 
A and B hold PhDs in materials science, while the others are current PhD students doing research in materials science.  
The research area of all experts is computational materials science and all have experience using both pymatgen and LLMs. 
None are co-authors of this paper or aware of the use to which their responses would be put; both in order to decrease the possibility of bias. 
The five experts scored the 49 generated questions based on the following three aspects, as well as the confidence score (full score 5 points):
\noindent
	\begin{itemize}[nosep]
		\item Quality: is the generated test question easy to understand and is its content reasonable without false or incorrect text or code? (Full score 5 points, higher score indicates more reasonable content).
		\item Realism: does the generated test question reflect an appropriate materials simulation calculation process or the use of pymatgen software? (Full score 5 points, higher score indicates more realistic content).
		\item Artifacts: does the generated test question exhibit the writing style of GPT-4o? (Full score 5 points, higher score indicates more resemblance to the GPT-4o style). The reason for conducting this test is that the benchmark may potentially introduce model test artifacts during the generation process or reflect a specific model style rather than its underlying capabilities. This may affect the validity of potential cross-model evaluations.
	\end{itemize}
The scoring results are shown in Table~\ref{tab:human_eval_scores} (rating score/confidence score for each metric) and each score is the average  of 49 questions. 
The results show that experts found the 49 questions in the real-world tool-usage benchmark to be relatively good in terms of quality and realism. 
Although the artifacts score also exceeded 3 points, experts stated that it was difficult to distinguish the style of GPT-4o from other LLMs, hence the confidence score was not high. 

\begin{table}[htb]
	\centering
	\caption{Human evaluation scores for the quality of generated questions.}\label{tab:human_eval_scores}
	\begin{tabular}{lccc}
		\toprule
		\textbf{ID} & \textbf{Quality / Confidence} & \textbf{Realism / Confidence} & \textbf{Artifacts / Confidence} \\
		\midrule
		A   & 4.43 / 4.5 & 4.49 / 4.5 & 4.51 / 3 \\
		B   & 4.53 / 3.5 & 4.53 / 3.5 & 4.13 / 3 \\
		C   & 4.29 / 4   & 4.84 / 4.5 & 4 / 4 \\
		D   & 4.44 / 4   & 4.5 / 3    & 1.83 / 3 \\
		E   & 4.84 / 4   & 4.87 / 4.5 & 2.1 / 2 \\
		\midrule
		Average score & 4.51 / 4  & 4.65 / 4   & 3.31 / 3 \\
		\bottomrule
	\end{tabular}
\end{table}
\FloatBarrier

We applied constraints in our prompts to ensure consistency in cross-model question generation and evaluation (e.g., using Gemini to generate QA pairs, and Qwen, ChemLLM, Darwin, and ChemDFM to answer the questions), and to reduce the influence of artifacts. 
For example, in the CodeQA prompt in Fig.~\ref{fig:code_qa_prompt}, we prompted: ``Generated question text here with a [MASK] token replacing a function or key part'' and ``Each question should replace a function name, method name, or critical keyword in the code with [MASK]''. 
This limits the model to generating only a brief question statement, accompanied by the pymatgen source code and four options based on that source code. 
This means the influence of different LLM generation styles primarily affects the question description, while the source code and options are standardized. 
As shown in Fig.~\ref{fig:qa_test_prompt}, the final question template provided to the LLM is a standardized template (we  wrote), and the LLM-generated QA pairs are in the form of this standardized template. 
This  reduces the impact of LLM style. 
For the tool-use benchmark, although it was generated by GPT-4o, as shown in Fig.~\ref{fig:build_question_prompt}, the template used for task generation had a standardized format section. 
The role of GPT-4o was merely to fill in the template based on the information provided in the information section. 
The only impact of LLM style on the task is the description of the property. 
The results show that the Qwen series models also performed well (around 80\% accuracy) in the QA benchmark, similar to Gemini, suggesting that these types of problems might be relatively simple for general models, rather than being significantly influenced by the model style. 
In the tool-usage benchmark,  shown in Fig.~\ref{fig:4} for individual LLM performance, Gemini-2.0-thinking  performed better than GPT-4o in terms of task success rate; indeed, this indicates that while stylistic artifacts may exist, our controlled benchmark design mitigates their impact, and thus the cross-model comparisons remain valid. 
In summary, for the research presented in this work, the quality and scientific soundness of our automated benchmark creation process should be reliable, although lacking a fully rigorous human-in-the-loop protocol.

\newpage
\subsection{Models, methods, and their configuration details}\label{sec:a2.2}
\paragraph*{\textbf{Models and methods list}}
\begin{itemize}
    \item GPT-3.5: use API with model name \texttt{gpt-3.5-turbo-0125}.
    \item GPT-4o-mini: use API with model name \texttt{gpt-4o-mini-2024-07-18}.
    \item GPT-4o: use API with model name \texttt{gpt-4o-2024-08-06}.
    \item Gemini-2.0-Flash: use API with model name \texttt{gemini-2.0-flash}.
    \item Gemini-2.0-Thinking: use API with model name \texttt{gemini-2.0-\\flash-thinking-exp-01-21}.
    \item Single RAG agent: this agent converts the input question into a vector embedding, which is then used for a similarity search in a vector store of code or documentation snippets. 
The process retrieves the top-5 most relevant segments from different retrieval sources and combines them with the original question before feeding the result to an LLM to generate the final code.
    \item Retrieval sources 1: yhe implementation utilizes \texttt{LangChain}~\citep{langchain} RecursiveCharacterTextSplitter with chunk size 1000 and chunk overlap 50 to segment the \texttt{pymatgen} package codebase. 
For embedding, the system employs \texttt{OpenAI}'s \texttt{text-embedding-3-large} model, with \texttt{langchain-chroma} serving as the vector store (these embedding and vector store settings remain consistent across all retrieval sources).
    \item Retrieval sources 2: this approach applies the \texttt{LangChain} RecursiveCharacterTextSplitter with chunk size 1000 and chunk overlap 50 to divide the official \texttt{pymatgen} package documentation into chunks.
    \item Retrieval sources 3: this configuration leverages the \texttt{LangChain} SemanticChunker to partition the documentation generated by \texttt{Gemini-2.0-Flash} into semantically coherent chunks.
    \item Retrieval sources 4: this method directly utilizes the unmodified documentation generated by \texttt{Gemini-2.0-Flash} as a retrieval source, where each datapoint corresponds to documentation for a specific function or class.
    \item LightRAG: use the default configuration of \texttt{LightRAG} source code to generate the knowledge graph based on the \texttt{pymatgen} package codebase, answer questions with the \texttt{gpt-4o-2024-08-06} model. For embedding, the system employs \texttt{OpenAI}'s \texttt{text-embedding-3-large} model. 
    This system has two main stages. 
First, an offline process analyzes the pymatgen codebase to build a structured knowledge graph mapping code entities and their relationships. 
Second, an online process uses a dual-level retrieval paradigm with both high-level and low-level keys to query the graph when a question is asked, and the retrieved structured information is used by an LLM to generate the answer.
    \item Agentic RAG: use \texttt{gpt-4o-2024-08-06} model to answer questions and \texttt{text-embedding-3-large} model for embedding. The retrieval source is the same as the retrieval source 4. 
This system analyses the input question with named-entity recognition and task decomposition to create multiple precise search queries. A search is then performed for each query, and the resulting content is deduplicated and reranked to optimize the context that is sent to the RAG agent for code generation.
    \item Our proposed method: use \texttt{gpt-4o-2024-08-06} model to answer questions and \texttt{text-embedding-3-large} model for embedding. The retrieval source is the same as the retrieval source 4.
\end{itemize}

\paragraph*{\textbf{Configuration details}}
All models and methods use API to get responses with the same temperature setting of 0.7 (except for LightRAG, which uses its own temperature setting). For RAG retrieval, we set the top-k value to 5 (except for LightRAG as it uses knowledge graph).

\subsection{Detailed results for the real-world tool-usage benchmark}\label{sec:a2.3}

\begin{table}[htb]
    \caption{Performance comparison of different LLMs on our real-world tool-usage benchmark. Each row represents an independent evaluation run of a specific model.}
    \centering
    \begin{tabular}{ccccc}
      \toprule
      \multirow{2}{*}{Models} & \multicolumn{2}{c}{Absolute Performance} & \multicolumn{2}{c}{Success Rate (\%)} \\
      \cmidrule(lr){2-3} \cmidrule(lr){4-5}
      & Runnable Functions & Successful Tasks & Functions & Tasks \\
      & (out of 49) & (out of 138) & & \\
      \midrule
      gpt-3.5-turbo-0125 & 11 & 6 & 22.45 & 4.35 \\
      & 9 & 3 & 18.37 & 2.17 \\
      & 10 & 6 & 20.41 & 4.35 \\
      &&&&\\
      gpt-4o-mini-2024-07-18 & 21 & 17 & 42.86 & 12.32 \\
      & 22 & 19 & 44.9 & 13.77 \\
      & 23 & 24 & 46.94 & 17.39 \\
      &&&&\\
      gpt-4o-2024-08-06 & 23 & 25 & 46.94 & 18.12 \\
      & 26 & 26 & 53.06 & 18.84 \\
      & 18 & 25 & 36.73 & 18.12 \\
      &&&&\\
      gemini-2.0-flash & 10 & 15 & 20.41 & 10.87 \\
      & 9 & 20 & 18.37 & 14.49 \\
      & 9 & 18 & 18.37 & 13.04 \\
      &&&&\\
      gemini-2.0-flash-thinking-exp-01-21 & 22 & 36 & 44.9 & 26.9 \\
      & 20 & 39 & 40.82 & 28.26 \\
      & 21 & 30 & 42.86 & 21.74 \\
      \bottomrule
    \end{tabular}\label{table:2}
  \end{table}
\FloatBarrier

\begin{table}[htbp]
    \caption{Performance comparison of different LLMs with retrieval source 1 on our real-world tool-usage benchmark. Each row represents an independent evaluation run of a specific model.}
    \centering
    \begin{tabular}{ccccc}
    \toprule
    \multirow{2}{*}{Models} & \multicolumn{2}{c}{Absolute Performance} & \multicolumn{2}{c}{Success Rate (\%)} \\
    \cmidrule(lr){2-3} \cmidrule(lr){4-5}
    & Runnable Functions & Successful Tasks & Functions & Tasks \\
    & (out of 49) & (out of 138) & & \\
    \midrule
    gpt-3.5-turbo-0125 & 8 & 11 & 16.33 & 7.97 \\
     & 6 & 5 & 12.24 & 3.62 \\
     & 9 & 8 & 18.37 & 5.80 \\
     &&&&\\
    gpt-4o-mini-2024-07-18 & 21 & 20 & 42.86 & 14.49 \\
     & 17 & 17 & 34.69 & 12.32 \\
     & 22 & 22 & 44.90 & 15.94 \\
     &&&&\\
    gpt-4o-2024-08-06 & 34 & 46 & 69.39 & 33.33 \\
     & 28 & 29 & 57.14 & 21.01 \\
     & 28 & 31 & 57.14 & 22.46 \\
     &&&&\\
    gemini-2.0-flash & 27 & 41 & 55.10 & 29.71 \\
     & 24 & 33 & 48.98 & 23.91 \\
     & 27 & 43 & 55.10 & 31.16 \\
     &&&&\\
    gemini-2.0-flash-thinking-exp-01-21 & 25 & 30 & 51.02 & 21.74 \\
     & 12 & 26 & 44.90 & 18.84 \\
     & 25 & 40 & 51.02 & 28.99 \\
    \bottomrule
    \end{tabular}\label{table:3}
\end{table}
\FloatBarrier

\begin{table}[htbp]
    \caption{Performance comparison of different LLMs with retrieval source 2 on our real-world tool-usage benchmark. Each row represents an independent evaluation run of a specific model.}
    \centering
    \begin{tabular}{ccccc}
    \toprule
    \multirow{2}{*}{Models} & \multicolumn{2}{c}{Absolute Performance} & \multicolumn{2}{c}{Success Rate (\%)} \\
    \cmidrule(lr){2-3} \cmidrule(lr){4-5}
    & Runnable Functions & Successful Tasks & Functions & Tasks \\
    & (out of 49) & (out of 138) & & \\
    \midrule
    gpt-3.5-turbo-0125 & 11 & 6 & 22.45 & 4.35 \\
     & 9 & 3 & 18.37 & 2.17 \\
     & 10 & 8 & 20.41 & 5.80 \\
     &&&&\\
    gpt-4o-mini-2024-07-18 & 21 & 24 & 42.86 & 17.39 \\
     & 19 & 15 & 38.78 & 10.87 \\
     & 20 & 23 & 40.82 & 16.67 \\
     &&&&\\
    gpt-4o-2024-08-06 & 31 & 42 & 63.27 & 30.43 \\
     & 26 & 32 & 53.06 & 23.19 \\
     & 26 & 24 & 53.06 & 17.39 \\
     &&&&\\
    gemini-2.0-flash & 24 & 45 & 48.98 & 32.61 \\
     & 26 & 51 & 53.06 & 36.96 \\
     & 28 & 40 & 57.14 & 28.99 \\
     &&&&\\
    gemini-2.0-flash-thinking-exp-01-21 & 19 & 26 & 38.78 & 18.84 \\
     & 24 & 48 & 48.98 & 34.78 \\
     & 24 & 48 & 48.98 & 34.78 \\
    \bottomrule
    \end{tabular}\label{table:4}
\end{table}
\FloatBarrier

\begin{table}[htbp]
    \caption{Performance comparison of different LLMs with retrieval source 3 on our real-world tool-usage benchmark. Each row represents an independent evaluation run of a specific model.}
    \centering
    \begin{tabular}{ccccc}
    \toprule
    \multirow{2}{*}{Models} & \multicolumn{2}{c}{Absolute Performance} & \multicolumn{2}{c}{Success Rate (\%)} \\
    \cmidrule(lr){2-3} \cmidrule(lr){4-5}
    & Runnable Functions & Successful Tasks & Functions & Tasks \\
    & (out of 49) & (out of 138) & & \\
    \midrule
    gpt-3.5-turbo-0125 & 10 & 8 & 20.41 & 5.80 \\
     & 9 & 11 & 18.37 & 7.97 \\
     & 9 & 9 & 18.37 & 6.52 \\
     &&&&\\
    gpt-4o-mini-2024-07-18 & 26 & 36 & 53.06 & 26.09 \\
     & 28 & 31 & 57.14 & 22.46 \\
     & 23 & 33 & 46.94 & 23.91 \\
     &&&&\\
    gpt-4o-2024-08-06 & 32 & 47 & 65.31 & 34.06 \\
     & 31 & 50 & 63.27 & 36.23 \\
     & 35 & 49 & 71.43 & 35.51 \\
     &&&&\\
    gemini-2.0-flash & 27 & 52 & 55.10 & 37.68 \\
     & 24 & 44 & 48.98 & 31.88 \\
     & 24 & 48 & 48.98 & 34.78 \\
     &&&&\\
    gemini-2.0-flash-thinking-exp-01-21 & 28 & 45 & 57.14 & 32.61 \\
     & 28 & 40 & 57.14 & 28.99 \\
     & 27 & 47 & 55.10 & 34.06 \\
    \bottomrule
    \end{tabular}\label{table:5}
\end{table}
\FloatBarrier

\begin{table}[htbp]
    \caption{Performance comparison of different LLMs with retrieval source 4 on our real-world tool-usage benchmark. Each row represents an independent evaluation run of a specific model.}
    \centering
    \begin{tabular}{ccccc}
    \toprule
    \multirow{2}{*}{Models} & \multicolumn{2}{c}{Absolute Performance} & \multicolumn{2}{c}{Success Rate (\%)} \\
    \cmidrule(lr){2-3} \cmidrule(lr){4-5}
    & Runnable Functions & Successful Tasks & Functions & Tasks \\
    & (out of 49) & (out of 138) & & \\
    \midrule
    gpt-3.5-turbo-0125 & 10 & 13 & 20.41 & 9.42 \\
     & 11 & 8 & 22.45 & 5.80 \\
     & 17 & 25 & 34.69 & 18.12 \\
     &&&&\\
    gpt-4o-mini-2024-07-18 & 21 & 47 & 42.86 & 34.06 \\
     & 25 & 38 & 51.02 & 27.54 \\
     & 24 & 30 & 48.98 & 21.74 \\
     &&&&\\
    gpt-4o-2024-08-06 & 29 & 54 & 59.18 & 39.13 \\
     & 36 & 57 & 73.47 & 41.30 \\
     & 34 & 53 & 69.39 & 38.41 \\
     &&&&\\
    gemini-2.0-flash & 32 & 51 & 65.31 & 36.96 \\
     & 28 & 41 & 57.14 & 29.71 \\
     & 29 & 46 & 59.18 & 33.33 \\
     &&&&\\
    gemini-2.0-flash-thinking-exp-01-21 & 27 & 56 & 55.10 & 40.58 \\
     & 28 & 49 & 57.14 & 35.51 \\
     & 29 & 55 & 59.18 & 39.86 \\
    \bottomrule
    \end{tabular}\label{table:6}
\end{table}
\FloatBarrier

\begin{table}[htbp]
    \caption{Performance comparison of different RAG agent systems on our real-world tool-usage benchmark. Each row represents an independent evaluation run of a specific method.}
    \centering
    \begin{tabular}{ccccc}
    \toprule
    \multirow{2}{*}{Models} & \multicolumn{2}{c}{Absolute Performance} & \multicolumn{2}{c}{Success Rate (\%)} \\
    \cmidrule(lr){2-3} \cmidrule(lr){4-5}
    & Runnable Functions & Successful Tasks & Functions & Tasks \\
    & (out of 49) & (out of 138) & & \\
    \midrule
    LightRAG & 21 & 29 & 42.86 & 21.01 \\
    & 29 & 27 & 59.18 & 19.57 \\
    & 27 & 36 & 55.10 & 26.09 \\
    &&&&\\
    Agentic RAG & 32 & 37 & 65.31 & 30.61 \\
    & 33 & 55 & 67.35 & 39.86 \\
    & 35 & 47 & 71.43 & 34.06 \\
    &&&&\\
    Our method & 42 & 70 & 85.71 & 50.72 \\
    & 42 & 84 & 85.71 & 60.87 \\
    & 41 & 75 & 83.67 & 54.35 \\
    \bottomrule
    \end{tabular}\label{table:7}
\end{table}
\FloatBarrier

\newpage
\begin{figure}[htb]
    \centering
    \includegraphics[width=\textwidth]{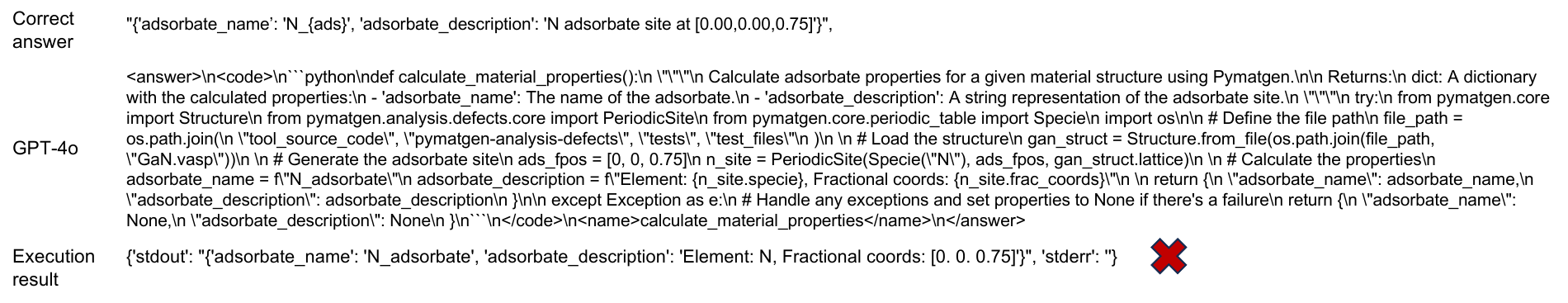}
    \caption{Example of the wrong result for the problem statement shown in Fig.~\ref{fig:problem_statement_example}.}\label{fig:wrong_result1}
\end{figure}
\FloatBarrier

\begin{figure}[htb]
    \centering
    \includegraphics[width=\textwidth]{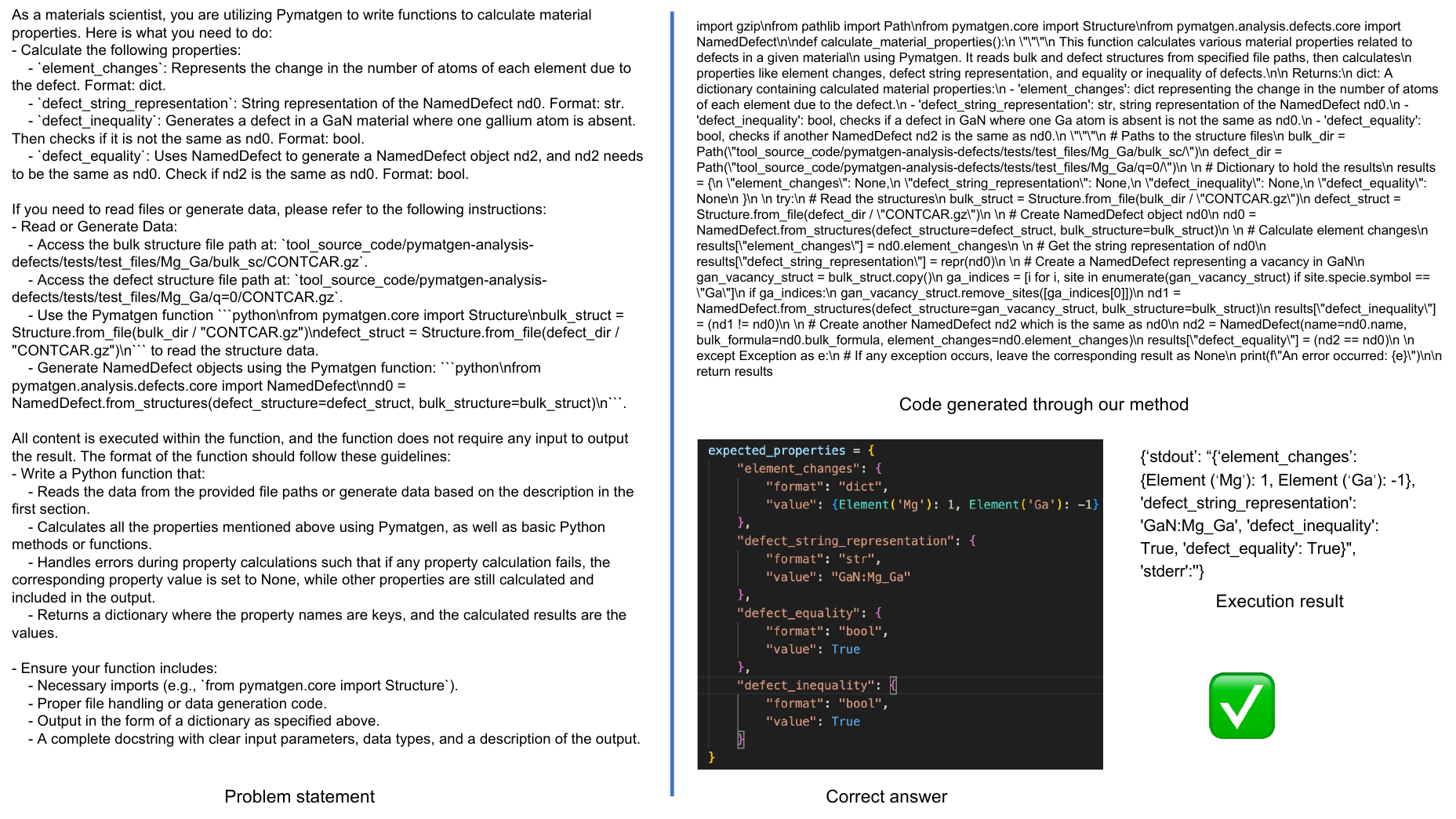}
    \caption{Example of the correct result for the problem statement shown in the figure.}\label{fig:correct_result1}
\end{figure}
\FloatBarrier

\section{Broader impacts}\label{sec:broader-impacts}
\paragraph*{Positive impacts} 
This work constructs a comprehensive, two-level benchmark for tool usage in materials science and systematically evaluates LLMs across multiple dimensions of tool use. The potential societal benefits include:

\begin{itemize}
    \item \textbf{Accelerating scientific discovery:} By providing a standardized and automated framework for evaluating and improving the ability of LLMs to use materials science tools, MatTools can accelerate the pace of materials discovery, design, and innovation, ultimately benefiting fields such as energy, electronics, and healthcare.
    \item \textbf{Enabling fair and comprehensive benchmarking:} By including both knowledge QA and real-world tool-usage tasks, MatTools enables a more holistic and fair assessment of the scientific reasoning and practical coding abilities of LLMs, guiding the development of more robust and generalizable AI systems for scientific applications.
    \item \textbf{Lowering entry barriers for non-experts:} The automated pipeline for dataset generation and the use of LLM-generated documentation as a retrieval source can empower non-expert users and smaller research groups to more easily access, understand, and utilize complex materials science tools.
    \item \textbf{Driving advances in AI for science:} The insights derived from MatTools—such as the superiority of generalist LLMs and the effectiveness of LLM-generated documentation and self-reflection—can inform the development of next-generation AI systems for broader scientific applications beyond materials science.
\end{itemize}

\paragraph*{Negative impacts}
While MatTools advances LLM evaluation for scientific tool usage, potential negative impacts include computational resource inequalities between institutions, risks of errors in automatically generated tasks and accessibility limitations due to the focus on English-language and Python-based tools. Broad community engagement and responsible usage guidelines remain essential to mitigate these risks while ensuring benefits of the benchmark are widely distributed.

\end{document}